\documentclass[pre,twocolumn,superscriptaddress,floatfix]{revtex4-1}

\usepackage{etex}
\usepackage{listings}
\usepackage[utf8]{inputenc}
\usepackage[text={7in,8.0in},centering]{geometry}
\usepackage{amsmath}
\usepackage{amssymb}
\usepackage{amsthm}
\usepackage{bbm}
\usepackage{tabularx}
\usepackage{url}
\usepackage{fancyhdr}
\usepackage{algpseudocode}
\usepackage{algorithm}
\usepackage{empheq}
\usepackage{color}
\usepackage{graphicx}
\usepackage{times}
\usepackage{hyperref}

\definecolor{lightyellow}{cmyk}{0,0,0.3,0}

\newcommand{\vc}[1]{{\pmb{#1}}}

\newcommand{\sv}{\vc{s}}

\newcommand{\tv}{\vc{t}}

\newcommand{\Gtilde}{\widetilde{G}}

\newcommand{\Wcal}{\mathcal{W}}
\newcommand{\Bcal}{\mathcal{B}}

\newcommand{\Ccal}{\mathcal{C}}

\newcommand{\Expect}[1]{\mathbb{E}_{#1}}

\newcommand{\instancename}{\texttt}

\begin{document}

\title{From Near to Eternity: Spin-glass planting, tiling puzzles, and
constraint satisfaction problems}

\author{Firas Hamze}
\affiliation{D-Wave Systems, Inc., 3033 Beta Avenue, Burnaby, British
Columbia, Canada, V5G 4M9}

\author{Darryl C. Jacob}
\affiliation{Department of Physics and Astronomy, Texas A\&M University,
College Station, Texas 77843-4242, USA}

\author{Andrew J. Ochoa}
\affiliation{Department of Physics and Astronomy, Texas A\&M University,
College Station, Texas 77843-4242, USA}

\author{Dilina Perera}
\affiliation{Department of Physics and Astronomy, Texas A\&M University,
College Station, Texas 77843-4242, USA}

\author{Wenlong Wang}
\affiliation{Department of Physics and Astronomy, Texas A\&M University,
College Station, Texas 77843-4242, USA}

\author{Helmut G.~Katzgraber}
\affiliation{Department of Physics and Astronomy, Texas A\&M University,
College Station, Texas 77843-4242, USA}
\affiliation{1QB Information Technologies, Vancouver, British
Columbia, Canada V6B 4W4}
\affiliation{Santa Fe Institute, 1399 Hyde Park Road, Santa Fe, New Mexico 
87501, USA}

\date{\today}

\begin{abstract}

  We present a methodology for generating Ising Hamiltonians of
  tunable complexity and with \emph{a priori} known ground states
  based on a decomposition of the model graph into edge-disjoint
  subgraphs. The idea is illustrated with a spin-glass model defined
  on a cubic lattice, where subproblems, whose couplers are restricted
  to the two values $\{-1,+1\}$, are specified on unit cubes and are
  parametrized by their local degeneracy.  The construction is shown
  to be equivalent to a type of three-dimensional constraint
  satisfaction problem known as the tiling puzzle. By varying the
  proportions of subproblem types, the Hamiltonian can span a dramatic
  range of typical computational complexity, from fairly easy to many
  orders of magnitude more difficult than prototypical bimodal and
  Gaussian spin glasses in three space dimensions. We corroborate this
  behavior via experiments with different algorithms and discuss
  generalizations and extensions to different types of graphs.

\end{abstract}


\maketitle

\section{Introduction}

Hard optimization problems are ubiquitous throughout the sciences and
engineering, and have consequently been the subject of intense
theoretical and practical study. While it is generally considered
highly unlikely that problems classified as NP-hard can be solved
efficiently for all members of their class, numerous algorithms have
been devised that either strive for an approximate solution (e.g.,
simulated annealing or stochastic local searches) or solve the
problems exactly in exponential time, but through mathematical insight
(e.g., branch-and-bound or branch-and-cut), the algorithms increase
the practically feasible sizes over what can be achieved using brute
force.  Evaluation of such algorithms often requires access to
benchmarking problems of various types; ideally their difficulty
should also be a controllable (or tunable) property.  Ideas of
generating test instances with planted solutions, that is, whose
optimizing values are known to the problem constructor, have been
explored in various fields for decades
\cite{bach:83,pilcher:87,pilcher:92}. A method of planting solutions
to hard random Boolean satisfiability (SAT) problems based on
statistical mechanics was first proposed in Ref.~\cite{barthel:02}.
In contrast, doing so for topologically structured problems is
considerably less charted territory, as the correspondence between
random SAT and the diluted spin glass disappears; thus replica
symmetry breaking analysis \cite{monasson:96} no longer applies. As
such, generating hard problems with a known solution for
nearest-neighbor Ising models on a hypercube is a relatively new field
of study.

\begin{figure}[tb]
    \includegraphics[width=\columnwidth]{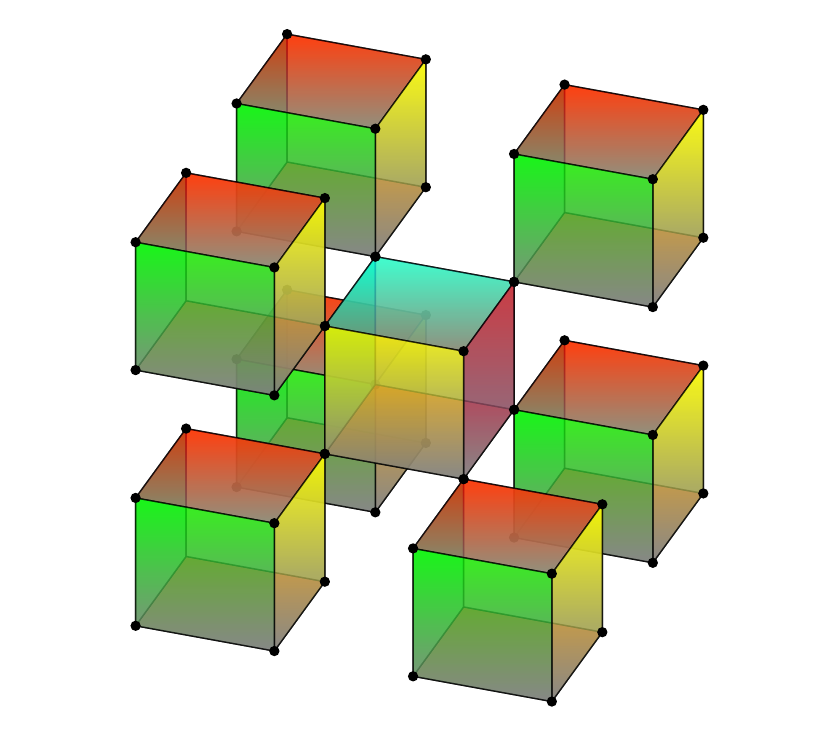}
    \caption{ Cutaway diagram showing a decomposition of the cubic
      lattice into edge-disjoint induced subgraphs, the colored unit
      cube cells. A cell in the right foreground is omitted for
      clarity.  The collection of vertices specifying each colored
      cube is defined as $\Ccal$. Under a periodic boundary, each spin
      appears in exactly two subgraphs. Ising subproblems with
      $J_{ij} \in \{-1,1\}$, i.e., requiring a single bit of
      precision, are specified for the cells such that one of each
      subproblem's ground states is consistent with its
      neighbors'. Over the full problem, the concatenation of these
      partial states forms the planted solution. As discussed in the
      text, the unit cell Hamiltonians' lack of shared edges is a
      crucial property enabling large, tunable changes in complexity
      while guaranteeing a known ground state.}
\label{fig:BCCPartition}
\end{figure}

While the task of creating such problems is of theoretical interest due
to their potential assistance in answering open questions about the
nature of spin glasses, the pace of research has been hastened in recent
years by practical motivations, in particular, the availability of
quantum annealing \cite{johnson:11,kadowaki:98} and related analog
devices (e.g., optical \cite{wang:13b}) physically implementing Ising
Hamiltonian minimization heuristics. Such Hamiltonians are usually
constrained by device manufacturing considerations to having
short-ranged interaction terms \cite{choi:08}; while more general
objectives can typically be encoded onto their native graphs, this may
require many system variables for each objective function variable,
limiting the problem sizes that can be studied. Hence, a way of encoding
topology-native problems is desirable.

In this work, we present a methodology for generating short-range Ising
problems with a known ground state. Our primary focus is on the
three-dimensional lattice with periodic boundary conditions, primarily
because the three-dimensional spin glass is a prototypical complex
system and of tremendous interest to condensed matter physicists, but
also because the regular structure over the lattice allows a description
of our key contributions in a natural and transparent manner. We stress,
however, that the method is adaptable to different graph topologies.

By carefully decomposing the underlying graphical structure and
selecting Ising subproblems over the resultant components, whose
spin-spin interactions are restricted to being bimodal, from classes
designed to have specific features influencing hardness, we obtain a
factor graph representation having the topology of a body-centered
cubic lattice (see Fig.~\ref{fig:BCCPartition}). Following a Voronoi
tessellation of this lattice, we note that the resultant problem is
equivalent to a specific type of three-dimensional tile matching
puzzle, a generalization of the two-dimensional edge-matching puzzle
\cite{demaine:07}. In their two-dimensional form, these puzzles are
constraint-satisfaction problems requiring placement of tiles from a
given set in permissible spatial locations so that patterns on the
tile edges match those of their neighbors. A well-known example is
Eternity II \cite{eternityII}, a to date unsolved problem for whose
solution a \$2 million prize was once offered; the prize deadline
expired in 2010.  Our construction yields polyhedral rather than
square puzzle pieces; in particular the units are truncated octahedra
whose facets are of various color patterns (see
Fig.~\ref{fig:octaPuzzle} below).  By varying the pattern properties
via manipulation of the underlying Ising subproblems, we quickly and
efficiently induce a tremendous variation in problem difficulty
observed over all examined heuristic algorithms; hundreds of problems
at any hardness level within our attainable range can be obtained in
seconds with modest computing resources. Hard problems within our
class are observed to be many orders of magnitude more difficult than
randomly generated spin-glass problems with bimodal disorder, i.e.,
where the spin-spin interactions are randomly drawn from $\{\pm 1\}$.
Despite our restriction to using problem parameters in $\{\pm 1\}$, we
can directly generate problems with a known solution that are vastly
more difficult than the Gaussian spin glass, which is itself known to
be considerably harder than the random bimodal class. We thus consider
our construction technique as giving rise to as yet unencountered
types of disordered system and discuss our results within the context
of hardness phase transitions for combinatorial problems.

The paper is structured as follows. In Sec.~\ref{sec:relWork} we present
an overview of recently-introduced planting techniques, followed by an
outline of the planting approach presented in this work in
Sec.~\ref{sec:probConst}.  Sections \ref{sec:spinGlassPuzzles} and
\ref{sec:phasesSat} discuss the relationship of the presented problems
to constraint satisfaction problems.  Section \ref{sec:experiments}
presents numerical experiments using different algorithms highlighting
the tunability of typical computational complexity for the problems with
planted solutions, followed by discussions, generalizations to other
graphs, and concluding remarks.

\section{Related Work}
\label{sec:relWork}

Our methodology addresses several shortcomings of recently proposed
techniques for solution planting in short-range Ising models. In
Ref.~\cite{hen:15a}, a construction method was presented in which
frustrated loops with known solution were employed as subproblem
units.  While the idea is appealing, the resultant problems typically
have coupler strengths spanning a large range, posing serious issues
for physical devices with precision limitations. An \emph{ad hoc}
limited-precision variant was introduced in Ref.~\cite{king:15}. Both
approaches, however, bear the more serious and systematic disadvantage
of tending to generate sets of problems whose members are not of
sufficient hardness; we contend that this is a consequence of adding
subproblem couplers. It is not hard to see that when partial
Hamiltonians having consistent ground states are added, couplers in
common which in the individual minima are either both satisfied or
both violated will increase in magnitude. In a ground state, however,
satisfied couplers are more common than violated ones, hence for any
given shared coupler, it is relatively likely that the resultant
(added) coupler tends to more strongly prefer alignment with the
shared ground state.  When this occurs over many edges, the problem's
degrees of freedom are ``softly'' constrained.  The effect is the
introduction of ``hints'' facilitating an optimization
algorithm's task.

In Ref.~\cite{marshall:16}, a method was presented for generating
difficult instances with bimodal couplers by iteratively changing the
couplers' signs to maximize the time required by a given solver to
reach the (hypothesized) ground state. This is a promising approach as
it both avoids the coupler addition issue and makes modest precision
demands; however, it generally does not allow the ground state to be
known with certainty and hence is difficult to scale to larger
systems.  Furthermore, the approach requires a Monte Carlo-like bond
moving algorithm that will likely not easily scale up to much larger
problem sizes.  The authors do discuss a variant enabling solution
planting, but as it again relies on adding couplers, the precision
issue returns and, they report, the gains in hardness are modest
compared to the mode not controlling the ground state. While the
authors provide a useful analysis of \emph{post hoc} empirical
correlates of hardness, for example the parallel tempering mixing
time, first proposed in Ref.~\cite{yucesoy:13}, the generation
procedure is ultimately a local search technique with respect to
solver computational time, and hence does not yield or use physical or
algorithmic insight at the problem level about what tunes
difficulty. In other words, one could not, merely by reference to the
resultant Hamiltonian, predict whether the problem is easy or
hard. Finally, to design problems based on the time to solution of a
given solver, assuming this will carry over to other optimization
techniques requires theoretical backing that is missing to date.

A recent paper \cite{wang:17} introduced the method of patch planting,
in which subgraphs with known solution are coupled to each other,
satisfying all interactions and thus planting an overall ground
state. An advantage is that low-precision and tunable problems can be
readily created for a given application domain. The authors report,
however, that the resultant problems are observed to be less difficult
than random problems, simply because frustrated loops typically do not
extend beyond the subgraph building blocks, making the ground state
less frustrated than in the fully random case. This problem can be
slightly alleviated by post processing mining of the data
\cite{wang:17}.

\section{Problem Construction}
\label{sec:probConst}

Consider a cubic lattice of linear size $L$ with periodic boundary
conditions. Let the underlying graph be $G = (V,E)$ with $V$ and $E$ its
vertex and edge sets, respectively. Define $U$ to be the vertices of $V$
whose coordinates $(i_x, i_y, i_z)$ consist of integers of the same
parity, for example, $(1,3,1)$ or $(0,2,4)$. For each $u \in U$, define
$C$ to be the vertices of the cubic unit cell implied when $u$ is
opposite to the vertex with coordinates $(i_x+1,i_y+1,i_z+1)$, where the
addition is modulo $L$ to account for the periodic boundary. Let $\Ccal$
be the collection of all such cell vertex sets. This construction is
partially illustrated in Fig. \ref{fig:BCCPartition}, where each
colored unit cell represents a subgraph $G[C]$ induced by a vertex set
$C \in \Ccal$ as defined. It should be clear that first, each vertex in
$V$ appears in two and only two unit cells and second, that the unit
cell subgraphs do not share any edges. In graph terminology, the family
$\Ccal$ partitions $G$ into edge-disjoint induced subgraphs.

We are concerned here with constructing Ising problems on the lattice
with zero field, i.e., whose Hamiltonians are of the form
\begin{equation}
  H(\sv) = \sum_{ij \in E} J_{ij}s_is_j .
\end{equation}
Due to the disjointness of the cell edge sets, we can regroup the
Hamiltonian into terms each dependent only on the couplings within a
unit cell in $\Ccal$,
\begin{equation}
  H(\sv) = \sum_{C \in \Ccal} H_C(\sv) ,
\end{equation}
where
\begin{equation}
 H_C(\sv) \triangleq \sum_{ij \in E[C]} J_{ij}s_i s_j .
\end{equation}
The terms $\{ H_C(\sv) \}$ are called the unit cell subproblem
Hamiltonians.  Specifying each subproblem’s Hamiltonian is sufficient to
imply a full Hamiltonian over the lattice. A straightforward consequence
of the relation between summation and minimization, exploited in
Ref.~\cite{hen:15a}, is that if the subproblems share the same
minimizing configuration $\sv^*$, then $H$ is also minimized at $\sv^*$.
This in turn suggests a natural construction procedure for a problem
$H$ with a known ground state. We note a seemingly small but in fact
deep and far-reaching difference between prior methodologies and what we
propose is that subproblem couplers are never added, avoiding the issue
discussed in Sec. \ref{sec:relWork}.

It may appear surprising at first that any interesting behavior can
result from linking such apparently simple unit-cell subproblems, but
this impression turns out to be quite false. Indeed, by restricting our
attention to simple classes of subproblems whose couplers belong to the
set $\{\pm 1\}$, in other words, representable with a single bit of
precision, we can effect dramatic changes in problem complexity, from
trivial to many orders of magnitude harder than, e.g., Gaussian spin
glasses.

\begin{figure}[tb]
    \includegraphics[width=\columnwidth]{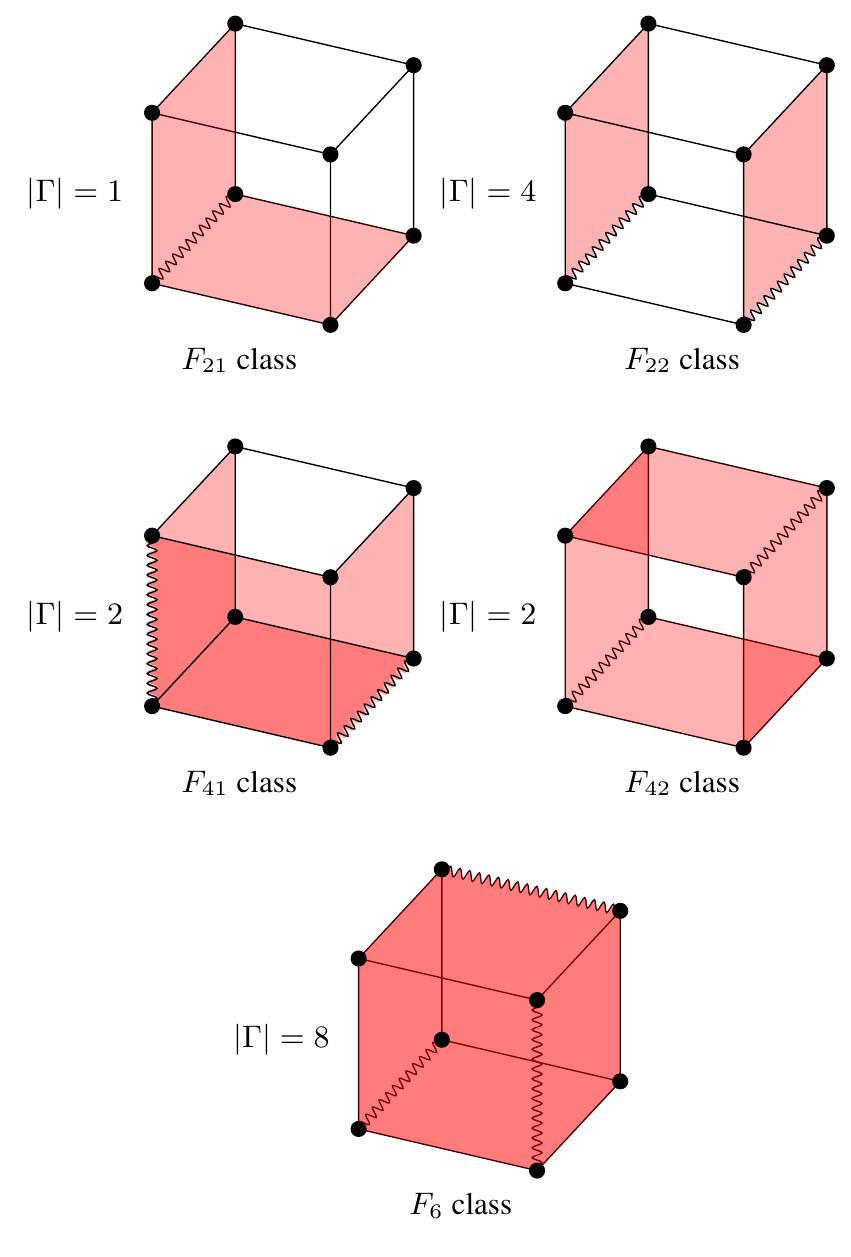}
    \caption{Class representatives of the three unit cell problem
      types grouped by the number of frustrated facets
      (shaded). Straight and wavy edges denote ferromagnetic and
      antiferromagnetic bonds, respectively. All problems have,
      generally among others, the ferromagnetic state
      $\sv^* = (+1,+1,\ldots,+1)$ as a Hamiltonian minimizer. The
      total number of ground states $|\Gamma|$ (up to ${\mathbb Z}_2$
      symmetry) for problems within each class is indicated beside
      each representative.}
\label{fig:FPProblems}
\end{figure}

\begin{figure}[tb]
    \includegraphics[width=\columnwidth]{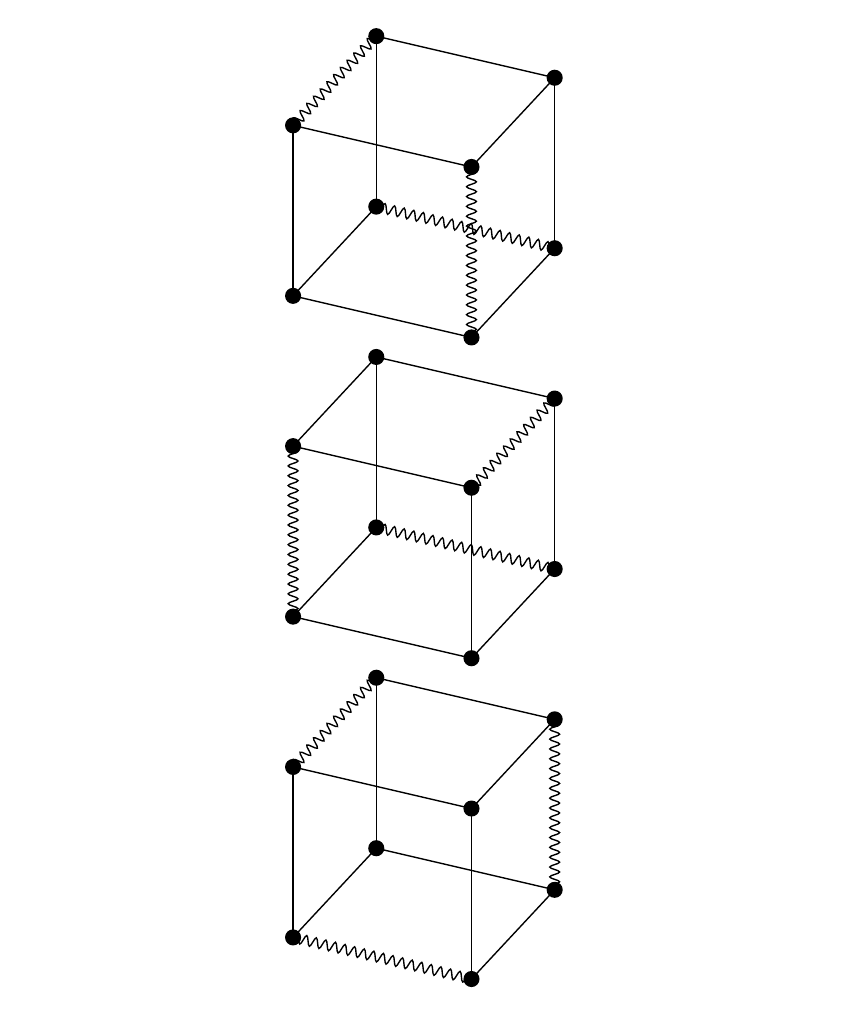}
\caption{Selected members of class $F_6$, equivalent under
    octahedral symmetry, that is, one of the $48$ compositions of
    rotation and reflection leaving an unmarked cube invariant.}
  \label{fig:F6Examples}
\end{figure}

Without loss of generality, we focus on planting the ferromagnetic
ground state, i.e., $\sv^* = (+1,+1,\ldots,+1)$ and its
${\mathbb Z}_2$ image, because once such a problem has been generated,
the structure of the construction procedure can be concealed by gauge
randomization from a would-be adversary seeking the solution.
Specifically, to translate to an arbitrarily chosen ground state
$\tv^*$, one would transform the initially determined couplings
$\{ J_{ij} \}$ via
\begin{equation}
J'_{ij} \gets J_{ij} t_i^*t_j^* .
\end{equation}
We thus turn our attention to the set of unit-cell Hamiltonians having
couplers of magnitude $1$ and ground state $\sv^*$. Here we disregard
the trivial subproblem comprised uniquely of ferromagnetic bonds;
those remaining can be naturally partitioned into three types
according to their number of frustrated facets. We call these types
$F_2$, $F_4$, and $F_6$, as they contain unit cells with two, four,
and six frustrated facets, respectively. These types can be further
partitioned into members equivalent under action of the cube graph
automorphism group.  More precisely, problems within a given
equivalence class can be arrived at from one another via
transformation from $O_h$, the $48$ octahedral symmetries comprised of
rotation and reflection leaving a (generic) cube invariant. It turns
out that $F_2$ and $F_4$ contain two such classes (the orbits
of $O_h$ acting on $F_2$ and $F_4$, called $F_{21}, F_{22}$ and
$F_{41}, F_{42}$) while all members of $F_6$ are equivalent under
$O_h$. Figure \ref{fig:FPProblems} illustrates an arbitrarily-chosen
set of equivalence class representatives for the three problems types,
while in Figure \ref{fig:F6Examples}, a few members of type $F_6$ are
shown.

\begin{figure}[tb]
    \includegraphics[width=\columnwidth]{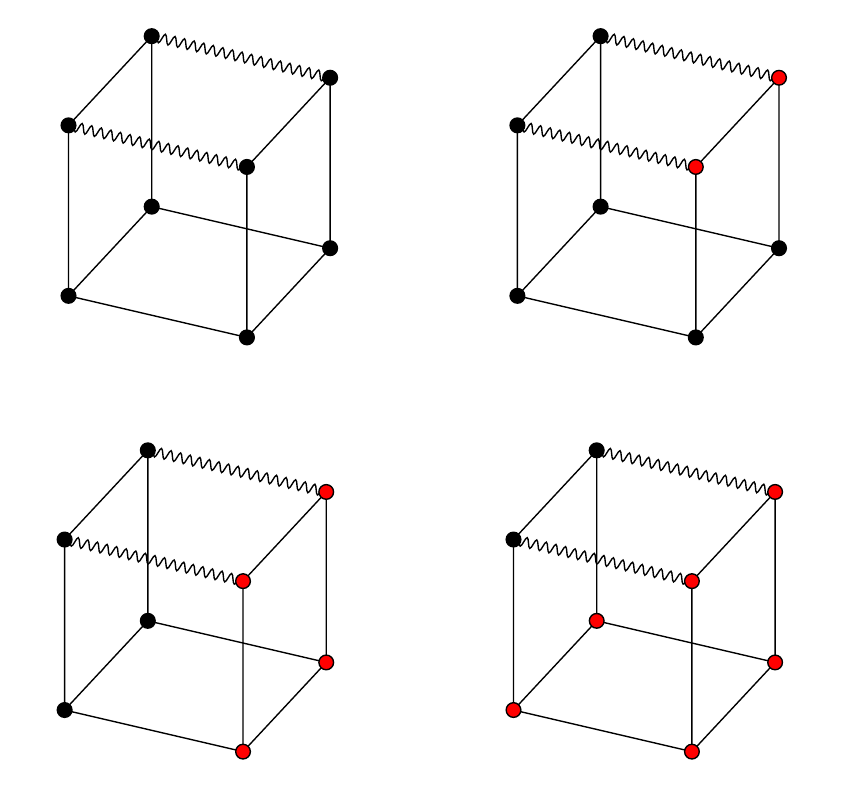}
  \caption{All subproblem ground states for a member of class
    $F_{22}$ (up to ${\mathbb Z}_2$ symmetry) where black (dark) and red
    (light) represent spins of opposite sign.}
  \label{fig:F22GSExamples}
\end{figure}

While the subproblems can all readily be verified to have $\sv^*$ as a
ground state, they clearly differ in the number of other minimizers
they possess.  Up to ${\mathbb Z}_2$ symmetry, members of $F_{21}$
have a unique ground state while those of $F_{22}$ have four. All
problems in $F_4$ and $F_6$ have two and eight global minima,
respectively. Figure \ref{fig:F22GSExamples} shows the four ground
states of a specific member of $F_{22}$. As we will see, this
variation in subproblem ground-state degeneracy turns out to exert a
crucial effect on the computational difficulty of the final planted
problem.

We are now ready to straightforwardly specify our problem construction
procedure. First, a probability distribution is chosen over the problem
classes $\{ F_{ij}\}$. A naive example, which in fact turns out to be
a poor choice for generating hard problems, is to sample them uniformly.
Next, a distribution over members of the classes is specified. In our
examples this was always uniform. The randomness in generating the
problems is thus decomposed into what we call subproblem disorder
(over the classes) and $O_h$ disorder (over the members), the
latter so called because it arises equivalently by sampling symmetries
of the octahedral group. Subproblems over the unit-cell set $\Ccal$ are
then sampled from the defined distributions.

Note that $O_h$ disorder is sufficient to induce considerable richness
in the types of final problems, even when the cells in $\Ccal$ are all
assigned from the same subproblem class.  For example, it may appear
at first that restricting all subproblems to the class $F_6$ (a regime
which turns out to be highly interesting) will result in a so-called
fully frustrated model \cite{villain:77}, i.e., where all chordless
cycles (or plaquettes) are frustrated in the final Hamiltonian,
because problems in $F_6$ are themselves fully frustrated.  That,
however, would be incorrect; it is easy to check that cycles forming
the facets of unit cubes not in $\Ccal$ may or may not be
frustrated depending on the random orientation of the bounding
$\Ccal$-cell problems.

It should be pointed out that while the method does ensure that the
planted state is in fact a ground state, as the couplers are selected
from the $\{\pm 1\}$ class, it is in principle possible for other
ground states to exist \cite{comment:free}. Our experimental
validation in Sec.~\ref{sec:experiments} is thus concerned with the
difficulty of finding any ground state.  Methods of removing the
degeneracy are certainly conceivable and are discussed briefly in
Sec.~\ref{sec:discussion}.

We proceed next to a further exploration of the problem structure,
discussing its connection to tiling puzzles.

\section{The puzzle of spin glasses}
\label{sec:spinGlassPuzzles}

The proposed problem generation formalism enables translation to a
specific type of geometrically appealing constraint-satisfaction
problem (CSP) called a tiling puzzle. While these problems are usually
defined over two-dimensional domains, we will see that our
three-dimensional technique yields a polyhedral generalization of an
edge-matching puzzle, an example of which is {\rm Eternity II}
\cite{eternityII}. The idea is to abstract the Ising problem into a
generic CSP representation known as a factor graph
\cite{kschischang:01}. While the Ising Hamiltonian itself can always
be interpreted as a factor graph, under our planting technique
additional structure is present, namely, the restriction to certain
subsets of states. Following construction of the CSP exploiting this
structure, the natural connection to a tiling puzzle is observed.

\begin{figure*}[tb]
    \includegraphics[width=0.62\columnwidth]{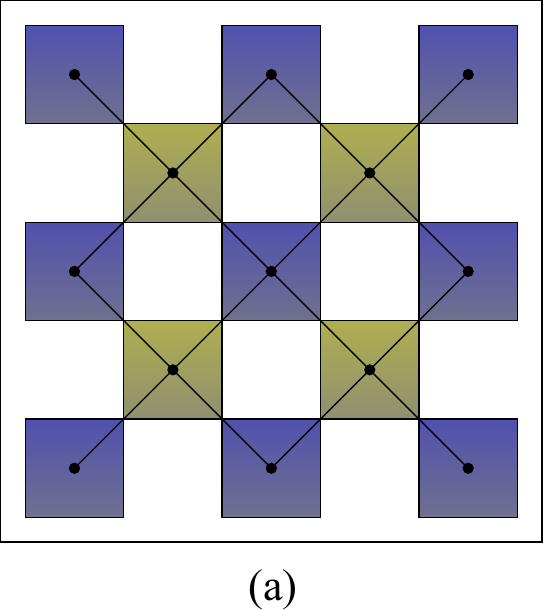}
    \hspace*{1em}
    \includegraphics[width=0.62\columnwidth]{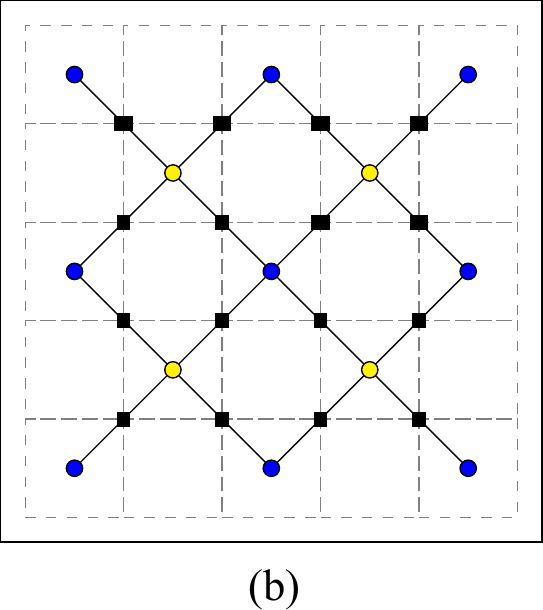}
    \hspace*{1em}
    \includegraphics[width=0.62\columnwidth]{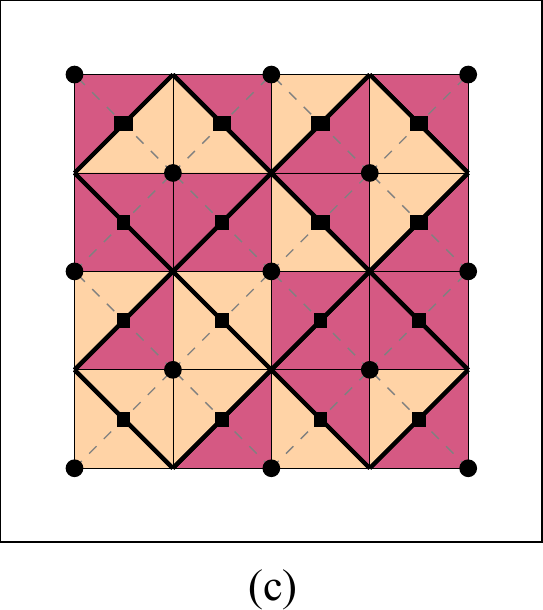}
    \caption{Steps in the derivation of a tile-matching puzzle from
      the unit-cell planting methodology, shown in two dimensions with
      a free boundary for clarity. Note that the two-dimensional cells
      induced by the decomposition are the unit cycle (plaquette)
      subgraphs arranged in a checkerboard pattern rather than the
      unit cubes shown in Fig.~\ref{fig:BCCPartition} in three
      dimensions. (a) The subproblem lattice $\Gtilde$ is constructed
      by placing a vertex at each subproblem cell's centroid. As
      discussed in the text, the vertices are 2-colored to yield a
      convenient representation of the puzzle. In three space
      dimensions, each yellow (light shading) unit cell contacts eight
      rather than four blue (dark shading) neighbors; the
      corresponding $\Gtilde$ is the body-centered cubic lattice. (b)
      Factor graph associated with $\Gtilde$, with factors shown as
      squares. Each vertex is involved in eight factors in three space
      dimensions. A Voronoi tessellation of $\Gtilde$ yields the set
      of tiling locations. In two space dimensions, these are the
      areas within the thick lines in panel (c). The factor graph
      domains, determined by the Ising subproblems, specify a set of
      allowable tile patterns at each location. The figure shows a
      hypothetical tile placed at each site. The factor constraints
      require adjacent colors to agree; hence, the central tile shown
      has one violation among its four neighbors. The puzzle for the
      three-dimensional tessellation is shown in
      Fig.~\ref{fig:octaPuzzle}.}
  \label{fig:tilingCSPSteps}
\end{figure*}

Figure \ref{fig:tilingCSPSteps} illustrates the necessary steps for the
mapping.  For clarity, a direct two-dimensional analog of the
three-dimensional case is shown, in which the unit cells induced by the
decomposition into $\Ccal$ are the cycles forming a checkerboard pattern
rather than unit cubes as displayed in Fig. \ref{fig:BCCPartition}.

We first consider the construction of a graph with respect to $\Ccal$,
which we call the subproblem lattice, denoted by $\Gtilde$.
Suppose a vertex is placed at the centroid of each unit cell. The
subproblem lattice is defined by considering two such vertices to be
adjacent if and only if their corresponding unit cells share a corner
vertex. Clearly, this lattice may be $2$-colored, that is, partitioned
into two subsets, none of whose members are neighbors. This in turn
implies a partitioning of the cells in $\Ccal$ into two disjoint sets
$\Wcal$ and $\Bcal$, whose elements are labeled with colors $W$ and $B$,
respectively. Figure \ref{fig:tilingCSPSteps}(a) demonstrates
the two-dimensional construction of $\Gtilde$ and its vertex coloring,
where yellow (light shading) and blue (dark shading) may represent
labels $W$ and $B$ or vice versa. In three space dimensions, each yellow
(light shading) cell is surrounded by eight blue (dark shading) cells
instead of four.

We next construct the factor graph CSP representation equivalent to the
problem of minimizing $H$.  The idea is to derive an objective
consisting of independent variables subject to equality constraints
along the edges of $\Gtilde$. First, each vertex in $\Gtilde$, mapping
to cell $C$ in the original lattice, is identified with a variable
$\sv_{C}^W$ or $\sv_{C}^B$, over eight Ising variables (four in two space
dimensions) depending on its label in the $2$-coloring. Because each
vertex of the initial lattice is a shared corner of two unit cells in
$\Ccal$ of opposite color, in the concatenation of the states
$\prod_{C\in \Wcal} \sv_C^W \prod_{C'\in \Bcal} \sv_{C'}^B$ variables
$s_i^W$ and $s_i^B$ will occur exactly once for each vertex $i \in V$.
The domain of each subproblem lattice variable is simply the ensemble of
its subproblem ground states, the ground-state set $\Gamma_C$, so that
the full configuration space of the new problem is the Cartesian product
\begin{equation}
\Gamma^{W,B}\triangleq \prod_{C\in \Wcal} \Gamma_C \prod_{C'\in
  \Bcal} \Gamma_{C'} .
\end{equation}
To be a feasible solution to the original problem however, equality of
neighboring variables must be enforced, that is, $s_i^W = s_i ^B$
$\forall$ $i$. If we define the problem
\begin{equation}
  \max_{\sv^{W,B} \in \Gamma^{W,B}} \prod_{i \in V}
  \delta[ s_i^W,s_i^B ] ,
  \label{eq:tilingCSP}
\end{equation}
where $\delta(x,y) = 1$ if $x=y$ and zero otherwise, we readily see
that the solution is obtained for the overall planted ground state
$\sv^B = \sv^W= \sv^*$. To make the graphical connection, we note that
each $\delta$ function in Eq.~(\ref{eq:tilingCSP}) can be interpreted
as a factor over neighboring subproblem lattice variables (identified
with) $C, C'$, i.e.,
$\psi_{C,C'} ( \sv_C^W, \sv_{C'}^B ) \triangleq \delta[ s_i^W,s_i^B] $
with $i = C \cap C'$. The final maximization objective is then the
product of all factors over subproblem lattice variables
\begin{equation}
  f(\sv^{W,B}) = \prod_{C,C' \in \widetilde{E}} \psi_{C,C'}( \sv^W,\sv^B) ,
\end{equation}
where the product index refers to edges of $\Gtilde$ by the cells in
$\Ccal$ mapping to their endpoint vertices. The factor graph
corresponding to the two-dimensional lattice in
Fig.~\ref{fig:tilingCSPSteps} (a) appears in
Fig.~\ref{fig:tilingCSPSteps} (b), with variables represented by
circular vertices (whose colors identify their labels as $W$ or $B$) and
factors by the squares along the edges. In three dimensions, each
variable is involved in eight factors rather than four.

Consider now the subproblem lattice with its vertices lying at their
``natural'' points in Euclidean space, i.e., at the centroids of the
unit cells, rather than as a generic graph. A Voronoi tessellation
\cite{torquato:13} is a partitioning of the space surrounding the
vertices into convex polytopes, usually called tiles, but for reasons
that will soon be clear, we refer to as tiling locations such that all
points within a polytope are closest (typically in the $L_2$-norm
sense) to the enclosed vertex. It is easy to see in
Fig.~\ref{fig:tilingCSPSteps}(a) that in two dimensions, the
subproblem lattice forms a periodic pattern, sometimes called the
quincunx, of squares of length $2$, each with a central vertex. The
appropriate tessellation of $\Gtilde$ is a partitioning into tilted
square regions centered at each vertex. Neglecting for an instant the
significance of the colors, Fig.~\ref{fig:tilingCSPSteps}(c) shows
five full and eight partial tiling squares in the section of lattice
drawn. Returning to the CSP in which the task was to assign the
variables from their respective domains (ground-state sets) so that
factor constraints are satisfied, we now observe the equivalence to a
tile-matching puzzle. This is comprised of a board (the space
embedding $\Gtilde$), segmented into octagons (the cells in the
Voronoi tessellation), each of which is endowed with a repertoire of
tiles (the ground-state set associated with the location). The tile
faces could be in one of two colors (the spin values, not to be
confused with the $2$-coloring introduced to construct the CSP). The
puzzle task is to select the tiles so that no two adjacent tile faces
have a different color. If orange (light) and purple (dark) represent
spin values, say, $1$ and $-1$, respectively, then
Fig.~\ref{fig:tilingCSPSteps} (c) shows a hypothetical tiling
assignment. Here a violation occurs where the tile colors on both
sides of a factor differ.

\begin{figure}[tb]
    \includegraphics[width=\columnwidth]{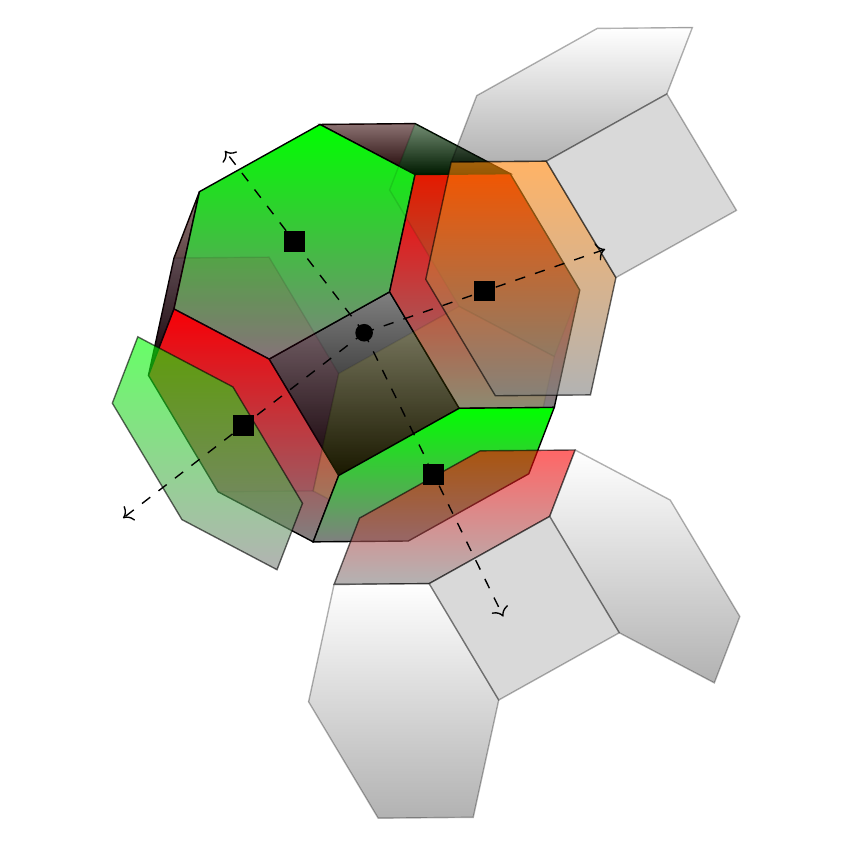}
    \caption{Tile-matching puzzle associated with three-dimensional
      unit-cell planting. The tile locations are truncated octahedra,
      the elements of a bitruncated cubic honeycomb, and in turn the
      Voronoi tessellation of the body-centered-cubic lattice
      $\Gtilde$. Color constraints are defined among neighboring
      hexagonal facets of the octahedra; the square facets are
      uninvolved. Four of the eight factors associated with a location
      are shown (the other four point into the page). In the example
      configuration shown, one constraint satisfaction and two
      violations occur.}
  \label{fig:octaPuzzle}
\end{figure}

In three dimensions, the puzzle is analogous, but with a few
twists. In a generalization of the two-dimensional case, the
subproblem lattice $\Gtilde$ will, when respecting Euclidean
coordinates, form a pattern of cubes of length $2$, each with a vertex
at its centroid. Such a formation is known as the body-centered-cubic
lattice. The corresponding Voronoi tessellation is the space-filling
bitruncated cubic honeycomb whose base unit is the truncated
octahedron \cite{conway:16,torquato:13}. Each of these polyhedral
cells, which now define the three-dimensional tiling locations,
consists of eight hexagonal facets, through which the the edges
of our CSP factor graph pass, and six square facets, defining the
boundaries between $\Gtilde$ vertices at distance $2$, but through
which no CSP factors pass. The ground-state sets associated with each
vertex of $\Gtilde$ thus map to coloring configurations on the
hexagonal facets of the enclosing Voronoi cell, defining a set of
tiles, and again the task is to place the tiles while meeting all
coloring constraints. The construction is illustrated in
Fig.~\ref{fig:octaPuzzle}.

While technically fitting the definition of a tile-matching puzzle, the
specific examples proposed here differ in a key way from conventional
puzzles such as Eternity II. In the latter, each tile from the
given set may be placed at more than one board location (in an allowable
orientation) while presently the ground-state set associated with each
site constrains the choices. This fact, along with the polynomial-time
solvability \cite{barahona:82} of the original ground-state problem in
two dimensions, could lead one to suspect that the puzzles we have
contrived may not be overly interesting. In three space dimensions,
however, solution by graph matching is no longer applicable, and as we
will see, the problems exhibit a rich range of behavior intimately tied
to the known properties of generic constraint satisfaction problems.

\section{The phases of satisfaction}
\label{sec:phasesSat}

Hardness phase transitions in combinatorial problems have been an
active area of study for several decades now \cite{hogg:96}. Of
particular interest is the emergence of difficulty divergencies in
random SAT problems as parameters guiding instance generation are
varied \cite{mezard:02}. While the tiling CSPs we have proposed here
can certainly be expressed as SAT problems \cite{comment:sub}, the
resultant formulas do not, however, follow the typical
assumptions of random SAT. Most notably, the variables are not free to
appear in any clause with some probability, but follow the
highly structured topological constraints of the puzzle.  Of course,
the subproblem and $O_h$ disorder impose stochasticity over the factor
graph variable domains $\{\Gamma_C\}$, but this is a localized
randomness, violating key assumptions in the analysis of random SAT.
Nonetheless, we expect that commonly known facts and intuitions
regarding the relation between how constrained a problem is and its
empirical hardness will apply to our situation.

Because the tiling puzzle color factors are invariant to the
distribution over cell problems, constrainedness depends exclusively
on the features of the ground-state sets. As we alluded to earlier, a
dominant correlate of problem hardness is the average size of the
ground-state sets $\Expect{} |\Gamma|$. A highly constrained regime is
one in which the subproblems tend to have relatively small degeneracy
and vice versa for a loosely constrained one. A subproblem
demonstrating a maximal constraint level is any member of class
$F_{21}$ (with a unique ground state up to ${\mathbb Z}_2$), while one
constrained minimally, which was not considered in the experiments, is
the zero $J$ subproblem, in which all $256$ states are ground states.
Within the class of subproblems introduced in
Sec.~\ref{sec:probConst}, including members of $F_{21}$ and $F_{6}$
(with degeneracy $8$) tends to bias the final problem towards the
maximal and minimal extremes, respectively.

\begin{figure*}[tb]
    \includegraphics[width=\columnwidth]{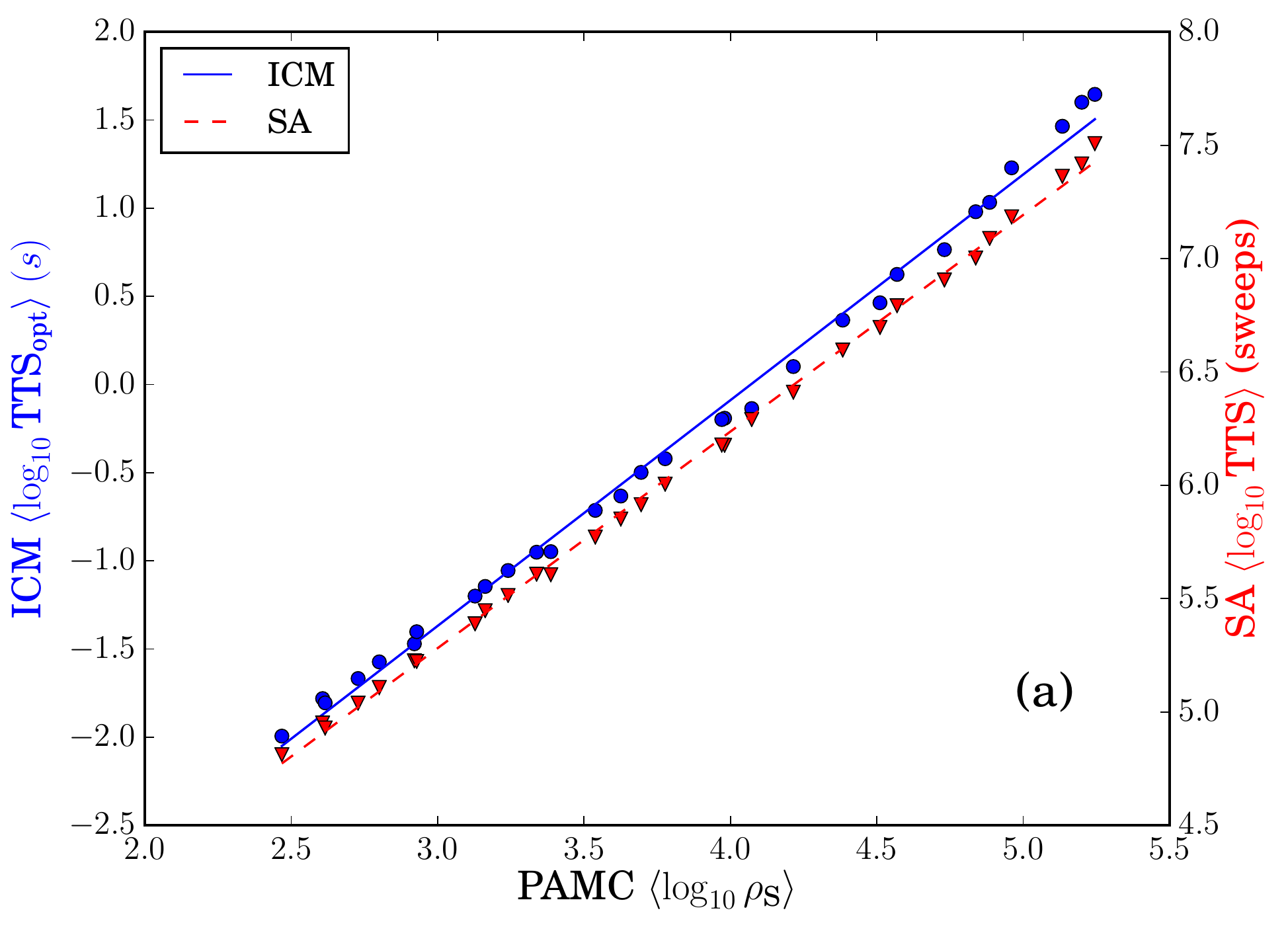}
    \includegraphics[width=\columnwidth]{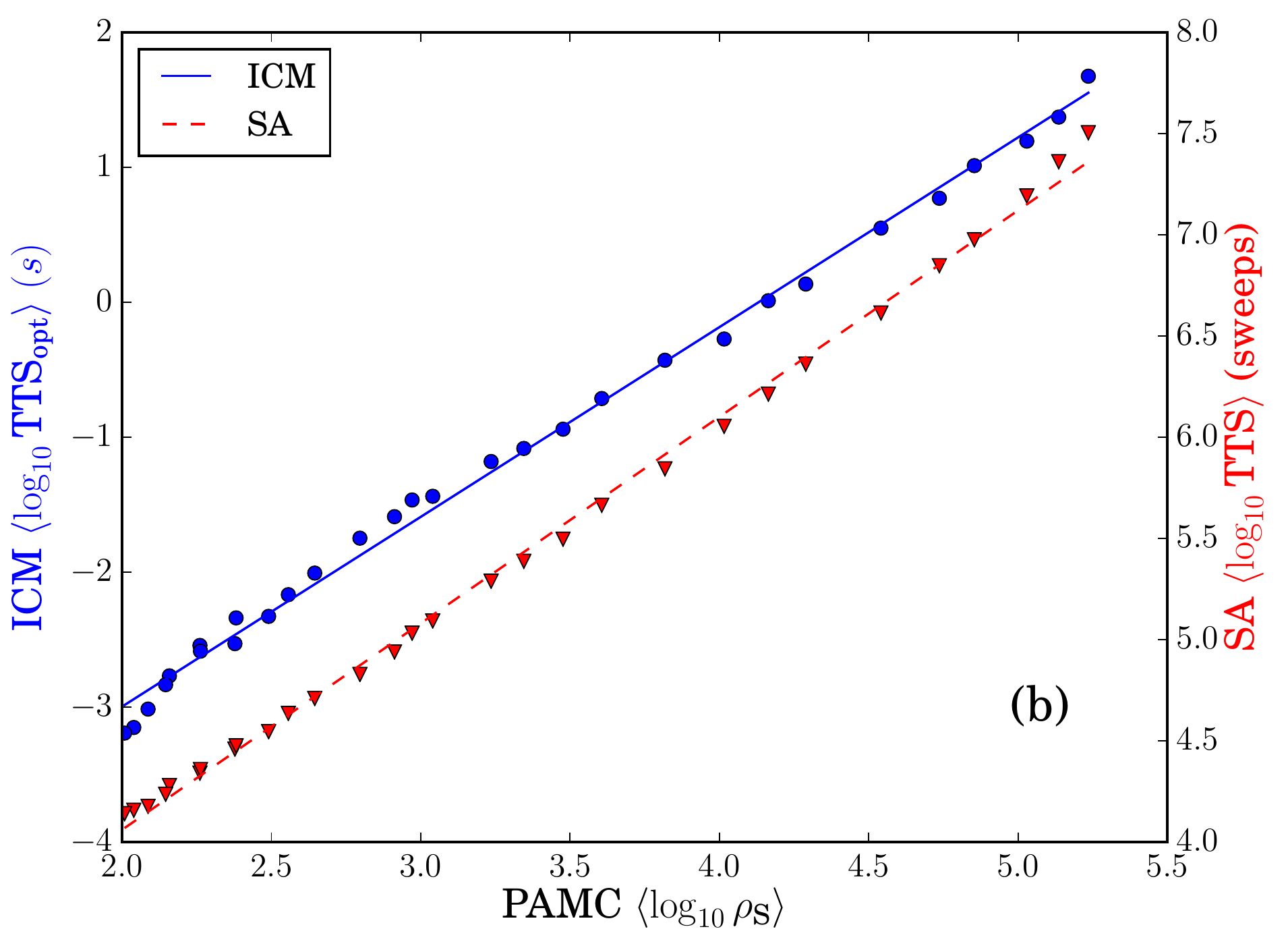}
    \caption{Simulated annealing $\langle \log_{10} {\rm TTS}\rangle$
      (right $y$-axes, red triangles) and parallel tempering with
      isoenergetic cluster moves
      $\langle \log_{10} {\rm TTS}_{\textrm{opt}}\rangle$ (left
      $y$-axes, blue circles) as a function of population annealing
      Monte Carlo $\langle \log_{10} \rho_{\rm S}\rangle$ for
      three-dimensional problem classes (a) \instancename{gallus\_26}
      and (b) \instancename{gallus\_46}.  The SA and ICM ${\rm TTS}$
      are measured in Monte Carlo sweeps and seconds,
      respectively. Within each problem class, every point in a given
      color (symbol) jointly shows the two corresponding measures for
      one of the $30$ subclasses described in the text (averages are
      computed over $200$ instances from each subclass). The data show
      that PAMC $\langle \log \rho_{\rm S}\rangle$ exhibits a strong
      linear correlation with the other two algorithms'
      $\log {\rm TTS}$ metrics, confirming its power as a measure of
      hardness and landscape roughness. Furthermore, as relative
      hardness measures for these problems, the three quantities
      studied here can be used more or less interchangeably. Error
      bars have been omitted for clarity.}
  \label{fig:ICMTTSsScatter}
\end{figure*}

By construction, all our problems are satisfiable, i.e., there is no
unsatisfiable (or overconstrained) regime as usually
understood. Clearly, the minimally constrained full problem is trivial
as each state is a solution to the original problem. In the tiling
puzzle representation, the factor constraints can be met independently
because the resultant states are always in the ground-state sets. At
the other extreme, highly constrained problems are also typically
easy: Because there are few local choices from the ground-state sets,
a tree search method will encounter relatively small branching factors
when traversing the state space, while for local search, the energy
landscape will be such that the algorithm can reliably locate the
ground state using simple strategies to escape from local
minima. These extremes strongly suggest that for some intermediate
level of mean ground-state set size, a peak in problem hardness will
occur. While it is not possible to increase $|\Gamma_C|$ past $8$ for
unit cells without introducing more complex Ising interactions (or
without trivially removing all $J_{ij}$), an interesting question,
probed experimentally in the following, is, given the proposed classes of
unit-cell subproblems ($F_2$, $F_4$, and $F_6$) and all possible
mixture distributions over the classes, did the hypothesized peak
occur for some interior point of the distribution set?  An affirmative
answer would suggest that the corresponding CSP class is among the
hardest of all tile-matching puzzles of the type we have introduced,
i.e., for which the locations' tile sets may be chosen arbitrarily,
not by constraining them to map to subproblem ground
states. Conversely, if the hardness maximum within the set of mixtures
occurs at the set boundary, it would point to the existence of more
difficult tiling puzzles not representable within our framework.

\section{Experiments}
\label{sec:experiments}

We now numerically study the typical complexity of certain subproblem
classes to illustrate their tunability. This section focuses on
three-dimensional cubic lattices, followed by a demonstration that the
approach works well also in two space dimensions
(Sec.~\ref{sec:chimera}). We conclude with a discussion of
generalizations to other graph types.

While we would ideally have performed a rigorous numerical study for a
discrete ``grid'' representing all subproblem types we have proposed,
this is computationally prohibitive. Consequently, we focus on two
illustrative regimes demonstrating interesting variations in problem
difficulty over three algorithms, and the role played by subproblem
degeneracy. The simulation results show that highly complex problems
with known ground state, and more difficult than conventional spin
glasses with both Gaussian and random bimodal couplings, are
accessible via our methodology, but they also strongly suggest that
the hardness maximum over the distribution set occurs at a boundary,
namely, when the distribution is a point mass concentrated on class
$F_6$. Perhaps unsurprisingly then, we conclude that tiling puzzles
equivalent to Ising problems constructed using the classes
$\{F_2,F_4,F_6\}$ are likely not the hardest among those where the
tile sets are arbitrarily specifiable.

All problems are defined on three-dimensional lattices of size $8\times
8\times 8$ with periodic boundary conditions. We consider two classes of
problems corresponding to one-dimensional slices of the problem mixture
parameter space. The first class, \instancename{gallus\_26}, allows
subproblems to belong solely to classes $F_6$ and $F_{22}$, while in the
second, \instancename{gallus\_46}, they are constrained to $F_6$ and
$F_{42}$. Both instance classes are parametrized by $p_6$, the
probability of selecting class $F_6$ instead of the alternative
\cite{comment:p6}. The classes each contained $200$ instances generated
at $30$ uniformly spaced values of $p_6 \in [0.8,1]$, for a total of
$6000$ instances per class. All subproblems are subject to uniform
$O_h$ disorder. The selected range of $p_6$ values is interesting as
within it, problems are more difficult than conventional spin glasses
with Gaussian or random bimodal couplings.  We note that the problems
continue to become easier as $p_6$ decreases.  Using the values of
$|\Gamma|$ shown in Fig.~\ref{fig:FPProblems}, the expected ground-state
set sizes in terms of $p_6$ are $\Expect{} |\Gamma|= 8p_6 + 4(1-p_6)$
for \instancename{gallus\_26} and $\Expect{} |\Gamma| = 8p_6 + 2(1-p_6)$
for \instancename{gallus\_46}.

\begin{figure*}[tb]
    \includegraphics[width=\columnwidth]{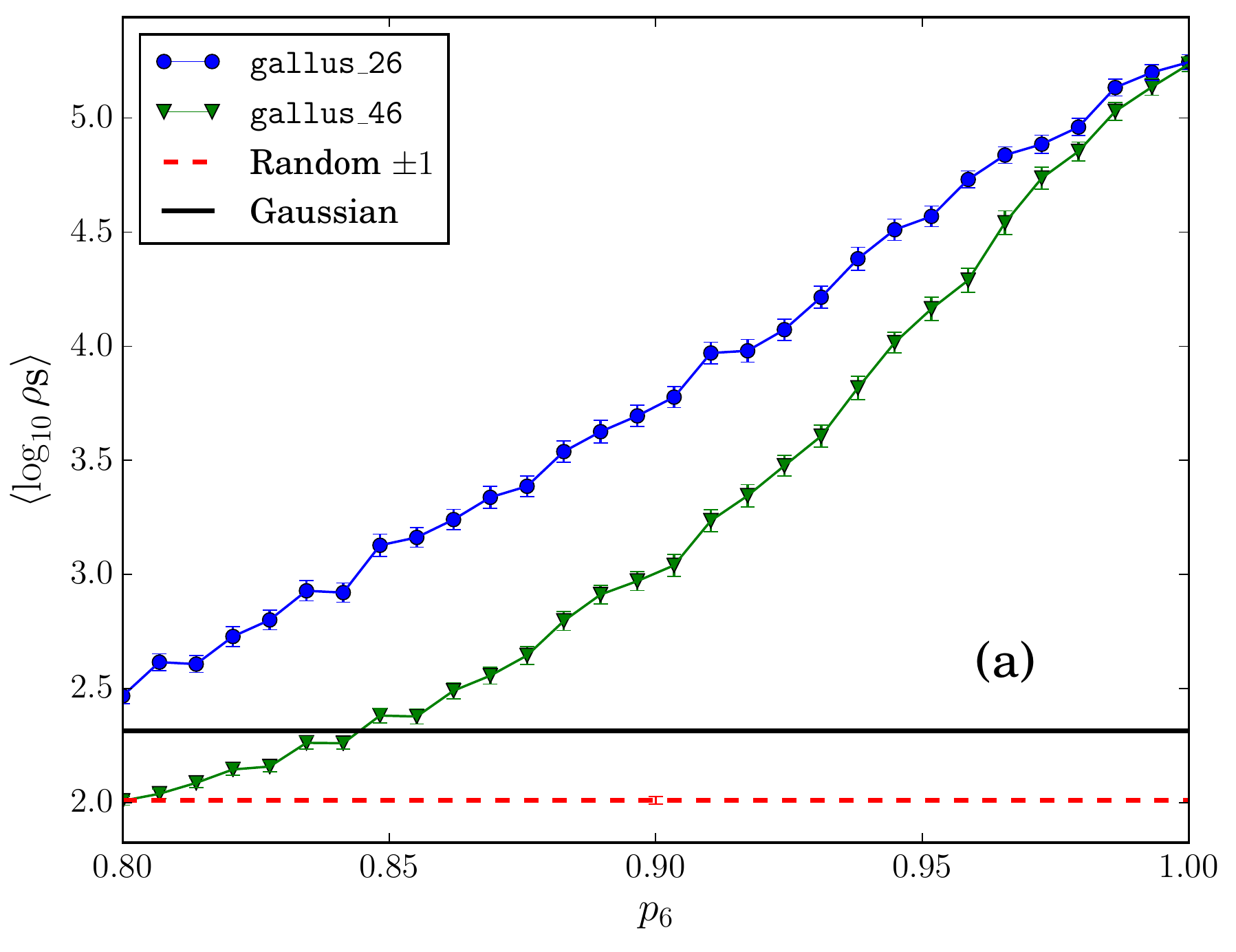}
    \includegraphics[width=\columnwidth]{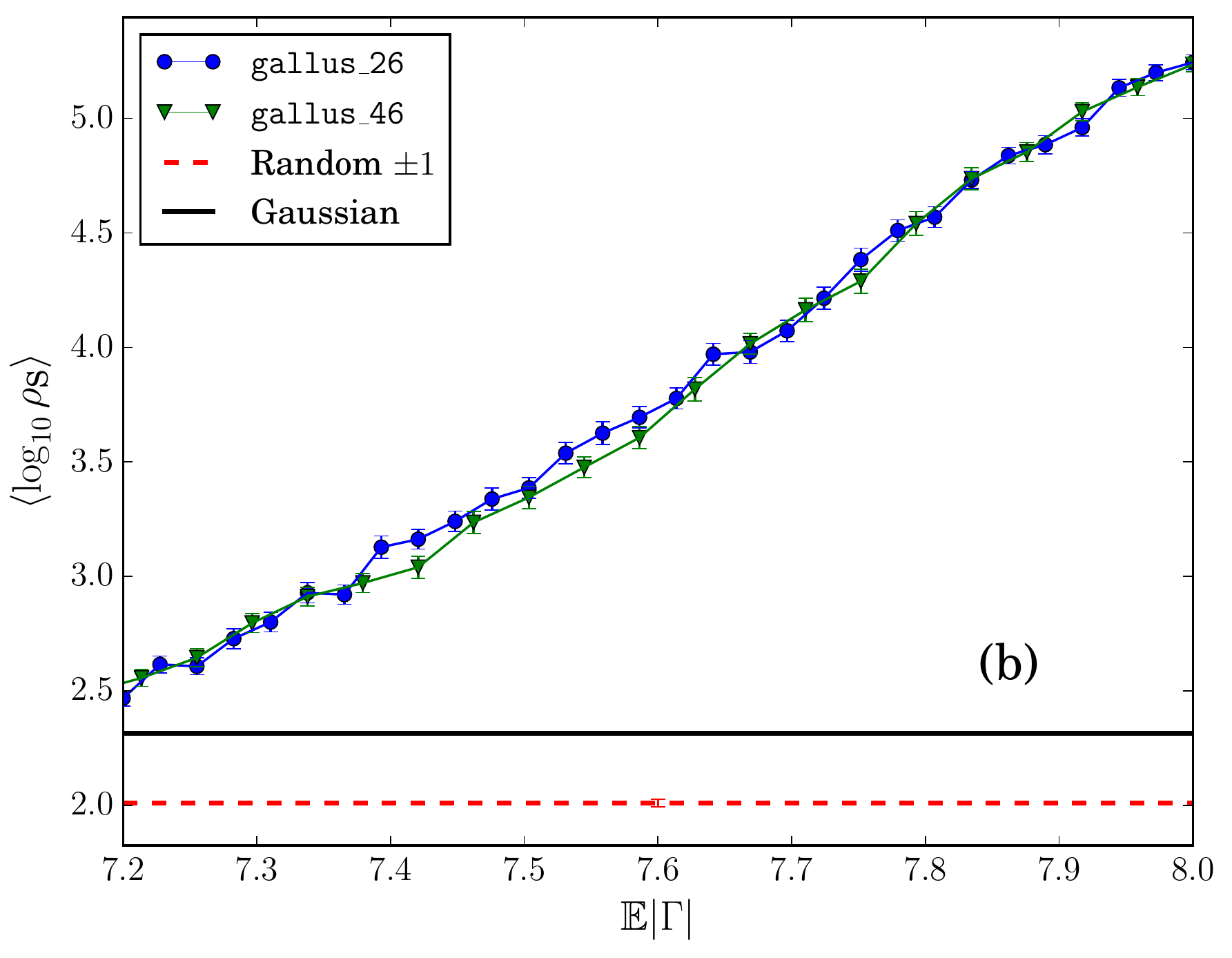}
    \caption{ Average log-entropic family size
      $\langle\log_{10} \rho_{\rm S}\rangle$ (population annealing
      Monte Carlo landscape roughness measure) for $L=8$ cubic lattice
      classes \instancename{gallus\_26} and
      \instancename{gallus\_46}. Each point corresponds to a subclass
      in which unit-cell subproblems are selected from $F_6$ with
      probability $p_6$, or else from $F_{22}$ for
      \instancename{gallus\_26} and from $F_{24}$ for
      \instancename{gallus\_46}. Averages are computed over $200$
      instances from each subclass. Plots of (a)
      $\langle \log_{10} \rho_{\rm S} \rangle$ against $p_6$ and (b)
      $\langle \log_{10} \rho_{\rm S} \rangle$ against
      $\Expect{}|\Gamma|$, the expected subproblem degeneracy within
      their range of overlap. For a given mixing probability $p_6$,
      (a) shows that \instancename{gallus\_26} is consistently more
      difficult than \instancename{gallus\_46} despite being less
      frustrated. On the other hand, (b) shows that instances
      from either class with a given subproblem degeneracy have very
      similar difficulty, suggesting that $\Expect{}|\Gamma|$
      considerably explains the variation in hardness irrespective of
      the specific underlying subproblem mixture or frustration level.
      In light of the proposed tiling puzzle interpretation, this is
      consistent with knowledge of CSP hardness.  For reference, note
      the $\langle \log_{10} \rho_{\rm S}\rangle$ values of
      equal-sized spin glasses with random $\pm 1$ and Gaussian
      interactions. For certain subclasses, the problems proposed here
      are far more difficult than the traditionally used cases.}
  \label{fig:PAlogRho}
\end{figure*}

Problem difficulty is assessed through performance of three different
algorithms designed for problems with rough energy landscapes. The
measures of difficulty are in strong agreement across the methods,
providing corroboration that the observed difficulty trends should
persist robustly across a fairly large class of heuristic algorithms,
including backtrack-based search \cite{hogg:96}.

Simulated annealing (SA) \cite{kirkpatrick:83} is the most basic
algorithm considered. We use the optimized implementation developed by
Isakov {\em et al}.~\cite{isakov:15} with $\beta_{\min} = 0.01$ and
$\beta_{\max} = 1$. While we would rather have used some form of
optimized time to solution (TTS) measure, which considers the best
tradeoff between simulation length and number of simulations, SA runs
were too time consuming on the hard problems to generate the requisite
runtime histograms reliably. Consequently, we selected a fixed run
length of $N_S = 8192$ sweeps for all problems with a single sweep per
temperature. Each problem is simulated $R = 10^6$ times independently.
The SA TTS is defined as the computational time required to find a
ground state with at least $99$\% probability, i.e., for each
instance, ${\rm TTS} = N_S \log(0.01)/\log(1 - p)$, where $p$ is the
fraction of successful runs out of the $R$ repetitions.

Furthermore, we have used a highly optimized implementation of
parallel tempering Monte Carlo with isoenergetic cluster moves (ICMs)
\cite{zhu:15b}, an adaptive hybrid parallel tempering (PT)
\cite{hukushima:96,geyer:91,moreno:03} cluster algorithm
\cite{houdayer:01}. Because the ICM is considerably more efficient
than SA, it allowed us to gather runtime statistics for each instance,
allowing optimized ${\rm TTS}$s to be computed. In contrast to SA,
total ICM computational time is not merely a function of overall Monte
Carlo sweeps, but includes the additional cost of constructing
random-sized clusters. Consequently, to track aggregated ICM
computational effort we record run times in seconds on hardware
dedicated entirely to the simulations. If $P(\tau)$ is the empirically
observed probability of finding the ground state in time $\tau$ or
less, the optimized time to solution is defined as
${\rm TTS}_{\textrm{opt}} = \min_{\tau} \tau
\log(0.01)/\log[1-P(\tau)]$.
We note, however, that despite the efficiency of the ICM, it commonly
failed to find the solution for the harder problems within the maximum
allowed $2^{24}$ total Monte Carlo sweeps, requiring $60-75$
min of real time, when computing the runtime histograms. For
difficult classes, these ``timeouts'' occurred at least half of the
time within the $100$ ICM repetitions used to tally the
histograms. Fortunately, we were nonetheless able to infer an
optimized ${\rm TTS}$ from the conditional histogram in which
solutions were found. We used $N_T=30$ temperatures spaced within
$T_{\max} = 3$ and $T_{\min} = 0.01$; see Ref.~\cite{zhu:15b} for
further details of the ICM.

The final set of tests have used the sequential Monte Carlo
\cite{delmoral:06} method known as population annealing Monte Carlo
(PAMC) \cite{hukushima:03,machta:10}. This algorithm is related to SA
but differs crucially in its usage of weight-based resampling, which
multiplies or eliminates members of a population according to the
ratios of their Boltzmann factors at adjacent temperatures,
maintaining thermal equilibrium at each step. Our simulations used
$N_T=201$ temperatures with $1/T = \beta\in[0.0,5]$ and $N_s=10$
sweeps per temperature, and a population size of $R=5\times10^5$
replicas. In Ref.~\cite{wang:15e}, a PAMC-derived index of landscape
roughness called the entropic family size $\rho_{\rm S}$ was
proposed. If $q_i$ is the fraction of replicas in the final population
descended from initial replica $i$, then
$\rho_{\rm S} = \lim_{R\to\infty}R/e^{h[q]}$, where
$h[q] = -\sum_{i}q_i\log q_i$. An energy landscape is thus deemed
rough if $h[q]$ is small, that is, if relatively few initial replicas
survive to the final distribution, yielding a large value of
$\rho_{\rm S}$. Note that $\rho_{\rm S}$ converges quickly in $R$ and
is easily estimated with finite populations. Thermal equilibration is
ensured by requiring $h[q] > \log 100 $; when this is not satisfied
for an instance, it is rerun with a larger population size.  The
entropic family size $\rho_{\rm S}$ is known to covary strongly with
the PT autocorrelation time \cite{wang:15e} and, as can be seen from
Fig.~\ref{fig:ICMTTSsScatter}, does so for the ${\rm TTS}$-based
hardness metrics considered here as well. Figure
\ref{fig:ICMTTSsScatter} shows, respectively, the dependence of both
SA and ICM $\langle \log_{10} {\rm TTS}\rangle$ measures on PAMC
$\langle \log_{10} \rho_{\rm S}\rangle$ for the $30$ subclasses of
\instancename{gallus\_26} [Fig.~\ref{fig:ICMTTSsScatter}(a)] and
\instancename{gallus\_46} [Fig.~\ref{fig:ICMTTSsScatter}(b)] studied,
where the averages $\langle \cdots \rangle$ at each point representing
a subclass are computed over its $200$ generated instances. The plots
clearly show a distinct near-linear dependence of both time-based
measures on the PAMC-defined value. A larger $\rho_{\rm S}$ on average
implies a longer time to solution with respect to both SA and ICM
algorithms, corroborating the former's power as an objective measure
of landscape roughness.

\begin{figure*}[tb]
    \includegraphics[width=\columnwidth]{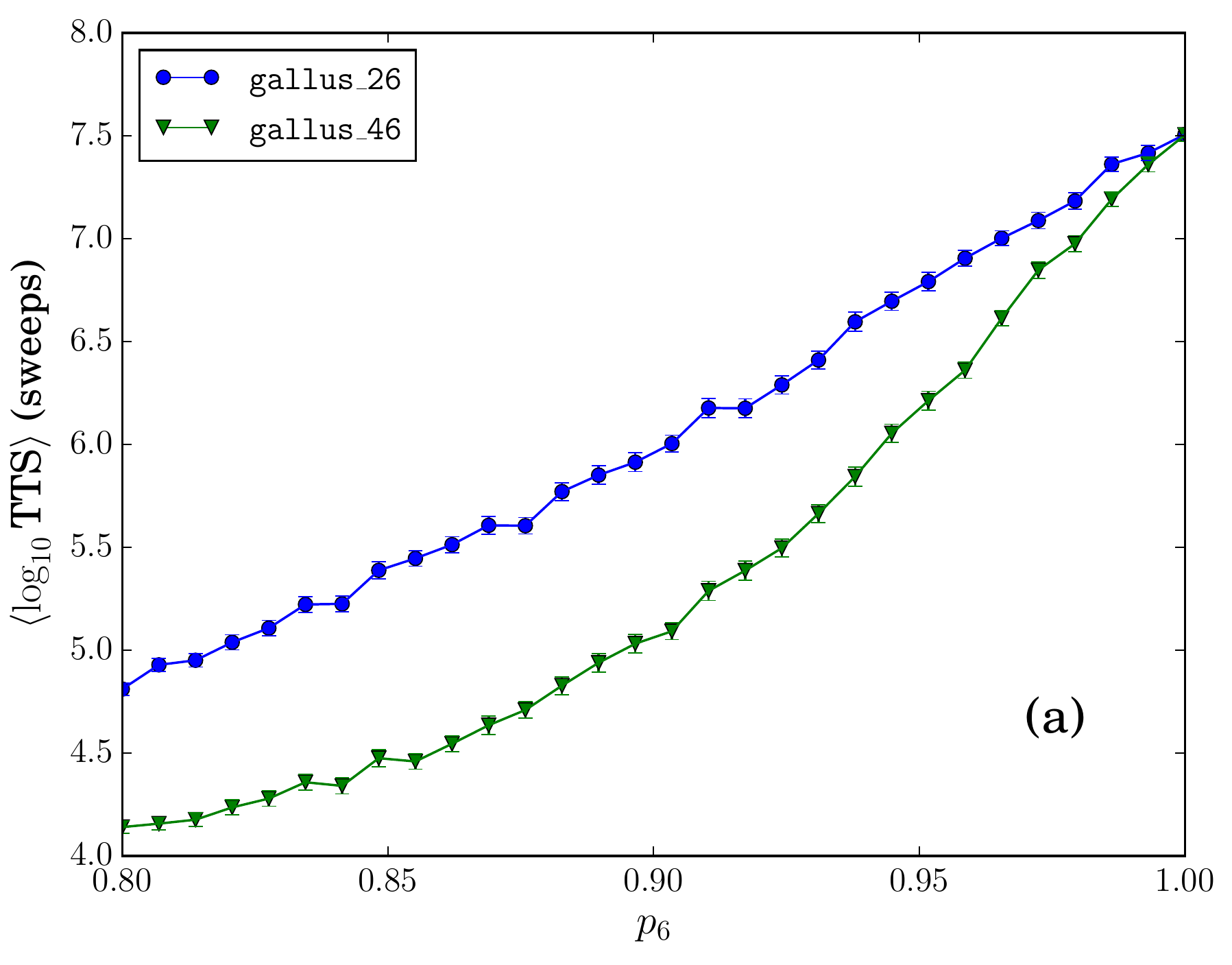}
    \includegraphics[width=\columnwidth]{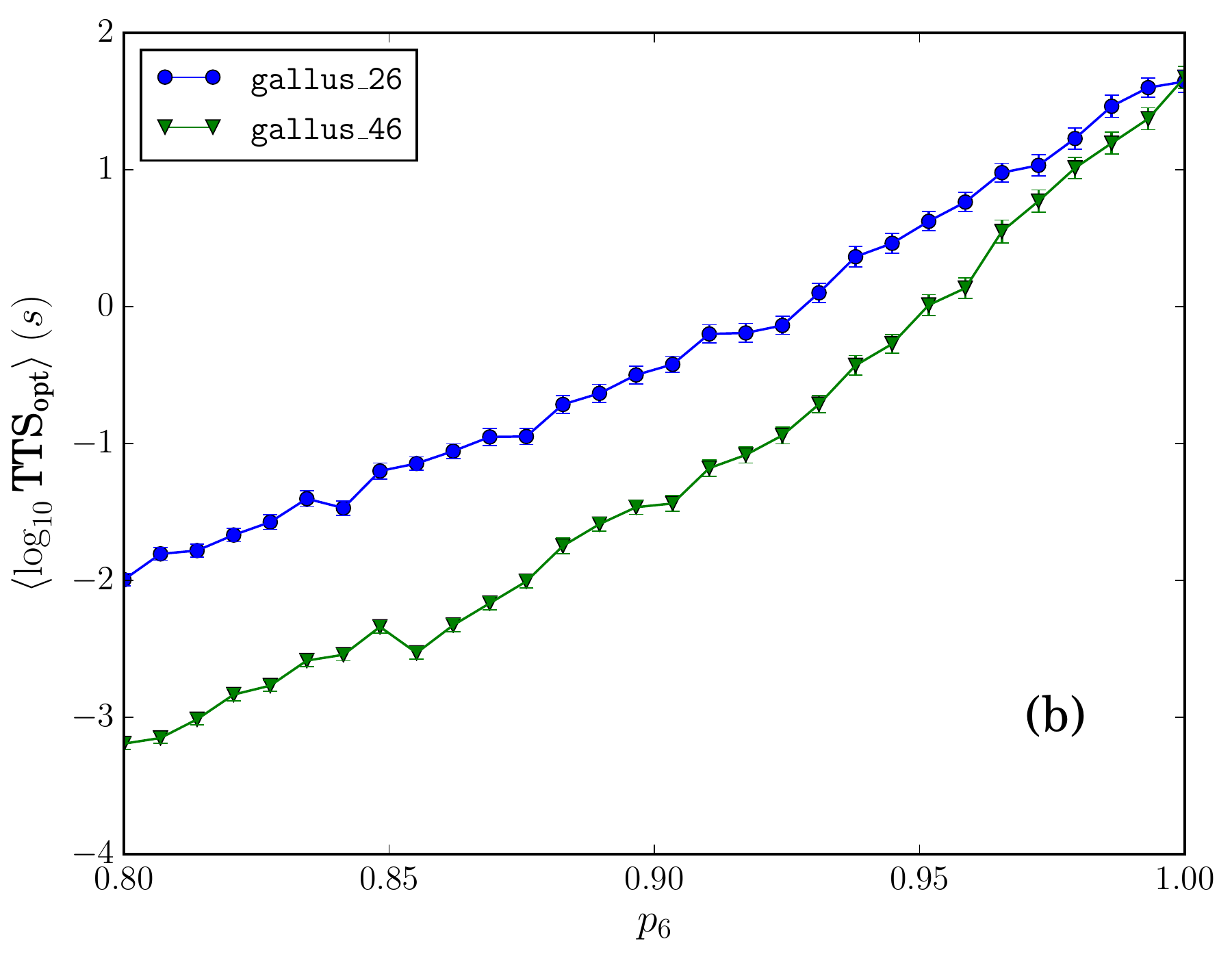}
    \caption{(a) SA $\langle \log_{10} {\rm TTS}\rangle$ and (b) ICM
      $\langle \log_{10} {\rm TTS}_{\textrm{opt}}\rangle$ (b) plotted
      against the parameter $p_6$ for $30$ subclasses of
      \instancename{gallus\_26} and \instancename{gallus\_46} defined
      in the text. The problems show the same relative difficulty
      trends with respect to both algorithms and in accordance with
      the PAMC results shown in Fig.~\ref{fig:PAlogRho} (a).}
  \label{fig:logTTS}
\end{figure*}

Results of the PAMC simulations are shown in Fig.~\ref{fig:PAlogRho}. We
observe a trend of increasing $\rho_{\rm S}$ (and hence complexity) for
both problem classes as the fraction of subproblems from $F_6$ increases
towards unity. For comparison, we display the mean $\log \rho_{\rm S}$
value of two prototypical problems with rough landscapes on the same
lattice, the random $J_{ij} \in \{\pm 1\}$ and Gaussian $J_{ij} \sim
N(0,1)$ [$N(0,1)$ a normal distribution with zero mean and variance one]
spin glass, computed using $1000$ and $5099$ instances of each type,
respectively. It is clear that problems in most of the examined
subclasses of \instancename{gallus\_26} and \instancename{gallus\_46}
are more difficult than both of these widely studied problem classes.
Indeed, for subclasses of \instancename{gallus\_26} corresponding to
$p_6 \in [0.91,1]$, problems are $\sim 2-3$ orders of magnitude
harder than bimodal spin glasses. Perhaps more surprisingly, they are
also $\sim 2-3$ orders more difficult than Gaussian spin glasses,
despite the latter possessing continuous-valued couplings while our
instances restrict couplers to $\{\pm 1\}$, which are believed to
generally be easier to minimize.

Setting $p_6=1$ yields the most complex problems of those considered. In
fact, we conjecture, based on less comprehensive simulations, that these
instances are the hardest among all those constructed with subproblem
classes in $\{F_2,F_4,F_6\}$. We plan to perform a comprehensive
analysis in the future.  For this hard class, the unit-cell subproblems,
deriving exclusively from $F_6$, have eight ground states each. Because
this hardness peak occurs at the boundary of the problem parameter
space, it seems plausible that one could instantiate yet more complex
three-dimensional tiling puzzles, where the locations were allowed more
than eight (times two) tile possibilities but not so many that the problem
becomes underconstrained and easy to solve.

At first, the greater pointwise difficulty shown in
Fig.~\ref{fig:PAlogRho} of an $\{F_6, F_{22}\}$ mixture over one made
of $\{F_6,F_{42}\}$ appears to contradict intuition that greater
frustration implies higher difficulty. After all, with four bounding
frustrated facets, a member of $F_{42}$ can be interpreted as more
frustrated than one of $F_{22}$, which has two. The story is somewhat
more subtle though: While frustration certainly plays a role in tuning
hardness in our problems, it appears to do so through its effect on
constraint level, namely, on the sizes of $\Gamma_C$, with $F_{22}$
inducing higher complexity than $F_{42}$ because its ground-state set
size is larger. This fact is displayed in Fig.~\ref{fig:PAlogRho}(b),
where $\log\rho_{\rm S}$ for the two classes is plotted against
$\Expect{}|\Gamma|$ instead of $p_6$. The graph shows that for a given
value of $\Expect{}|\Gamma|$, \instancename{gallus\_26} and
\instancename{gallus\_46} have a very similar roughness index,
implying that mean subproblem degeneracy accurately predicts
difficulty regardless of the underlying subproblem mixture.

For completeness, analogous plots displaying similar difficulty trends
for SAs $\log {\rm TTS}$ and ICMs $\log {\rm TTS}_{\textrm{opt}}$ are
shown in Fig.~\ref{fig:logTTS}, where they are again plotted against
$p_6$.

\begin{figure*}[tb]
    \includegraphics[width=\columnwidth]{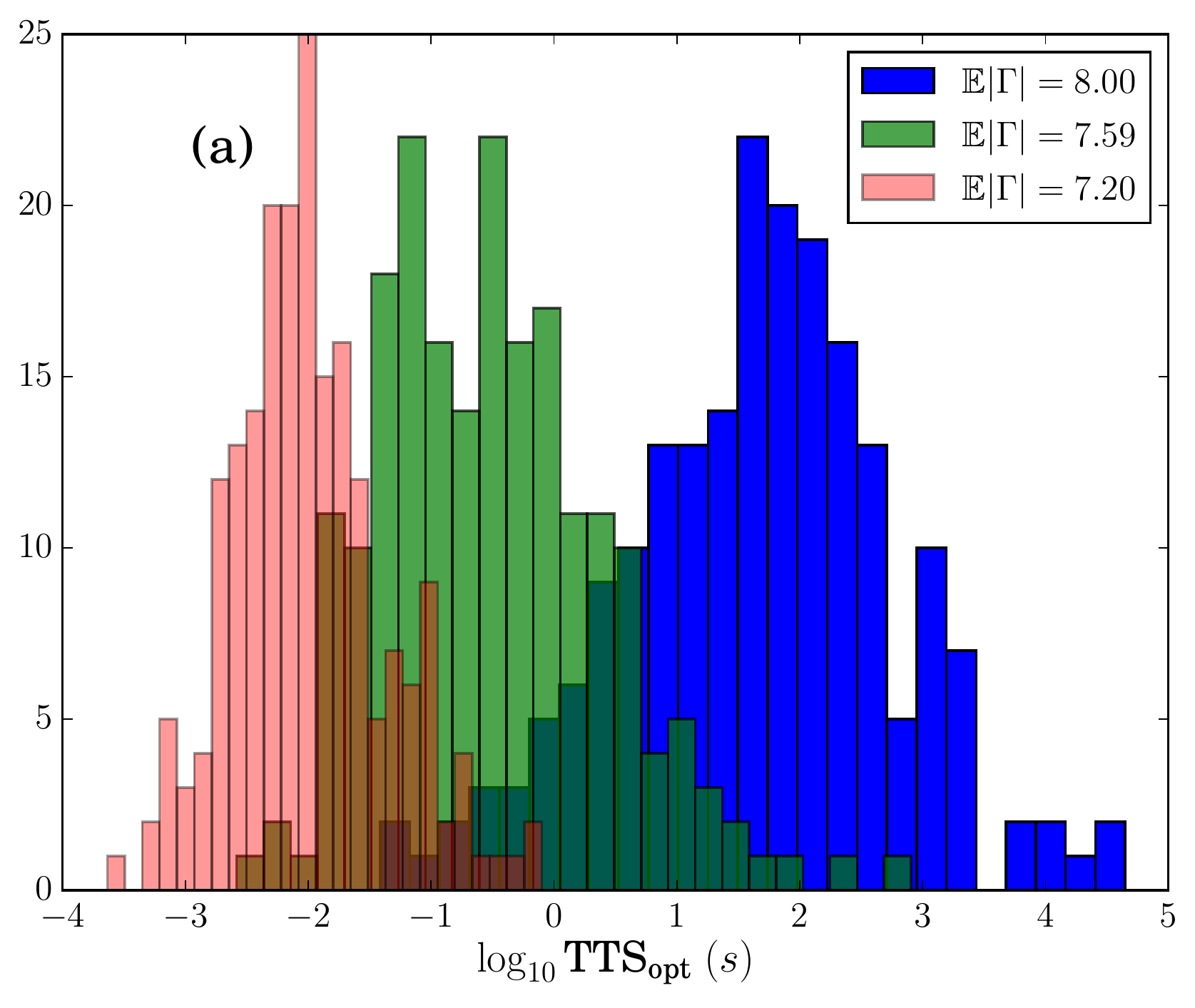}
    \includegraphics[width=\columnwidth]{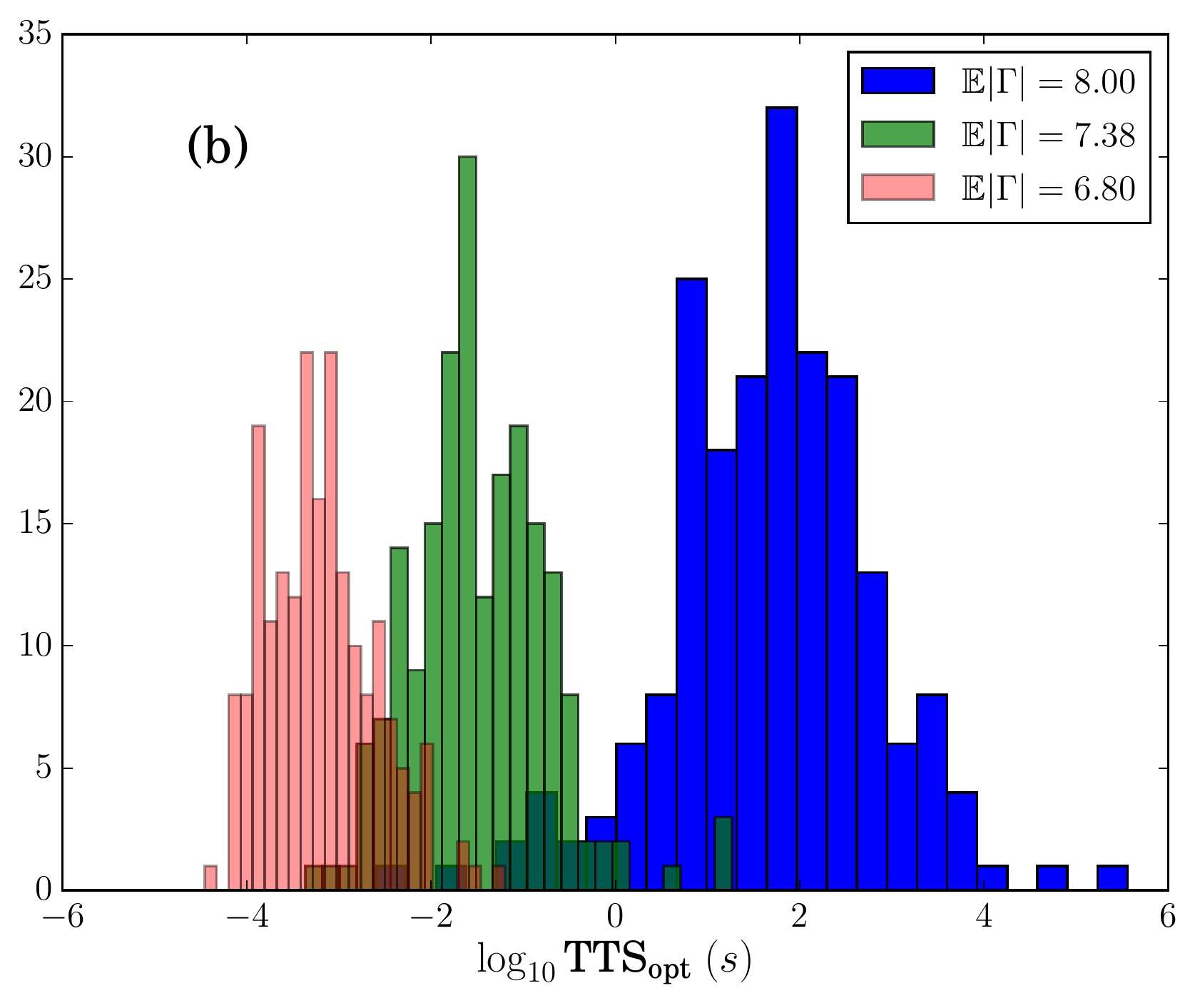}
  \caption{Histograms of (optimized) $\log_{10} {\rm
    TTS}_{\textrm{opt}}$ relative to isoenergetic cluster moves for
    classes (a) \instancename{gallus\_26} and
    (b) \instancename{gallus\_46} for three subclasses
    characterized here by mean subproblem degeneracy
    $\Expect{}|\Gamma|$; $200$ problems within each subclass were used
    to obtain the histograms. The leftward trend in both sets of
    histograms shows clearly that the problems become easier with
    decreasing $\Expect{}|\Gamma|$ and their shapes suggest that ${\rm
    TTS}_{\textrm{opt}}$ is log-normally distributed.}
  \label{fig:ICMlogTTSOptHistograms}
\end{figure*}

\begin{figure*}[tb]
    \includegraphics[width=\columnwidth]{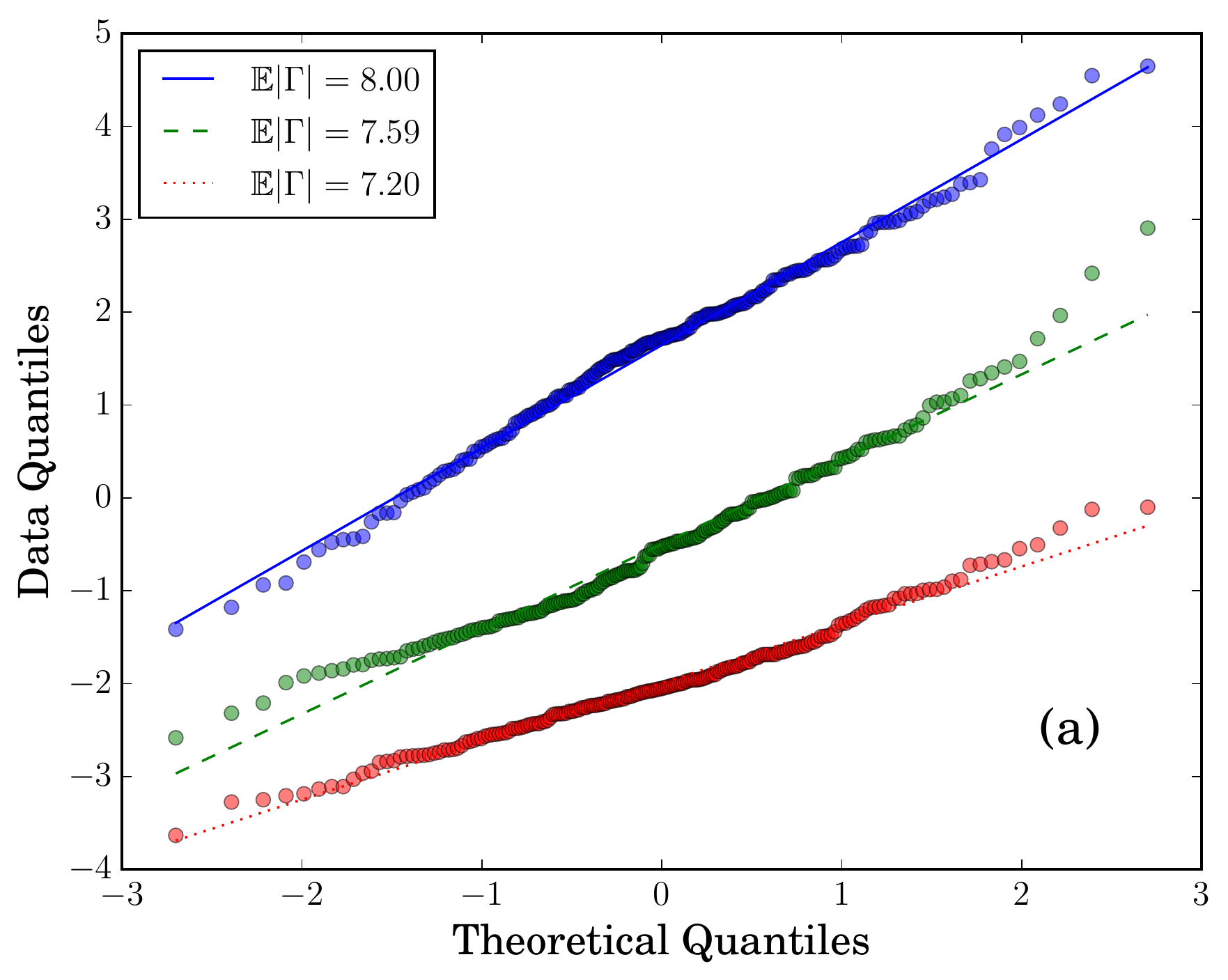}
    \includegraphics[width=\columnwidth]{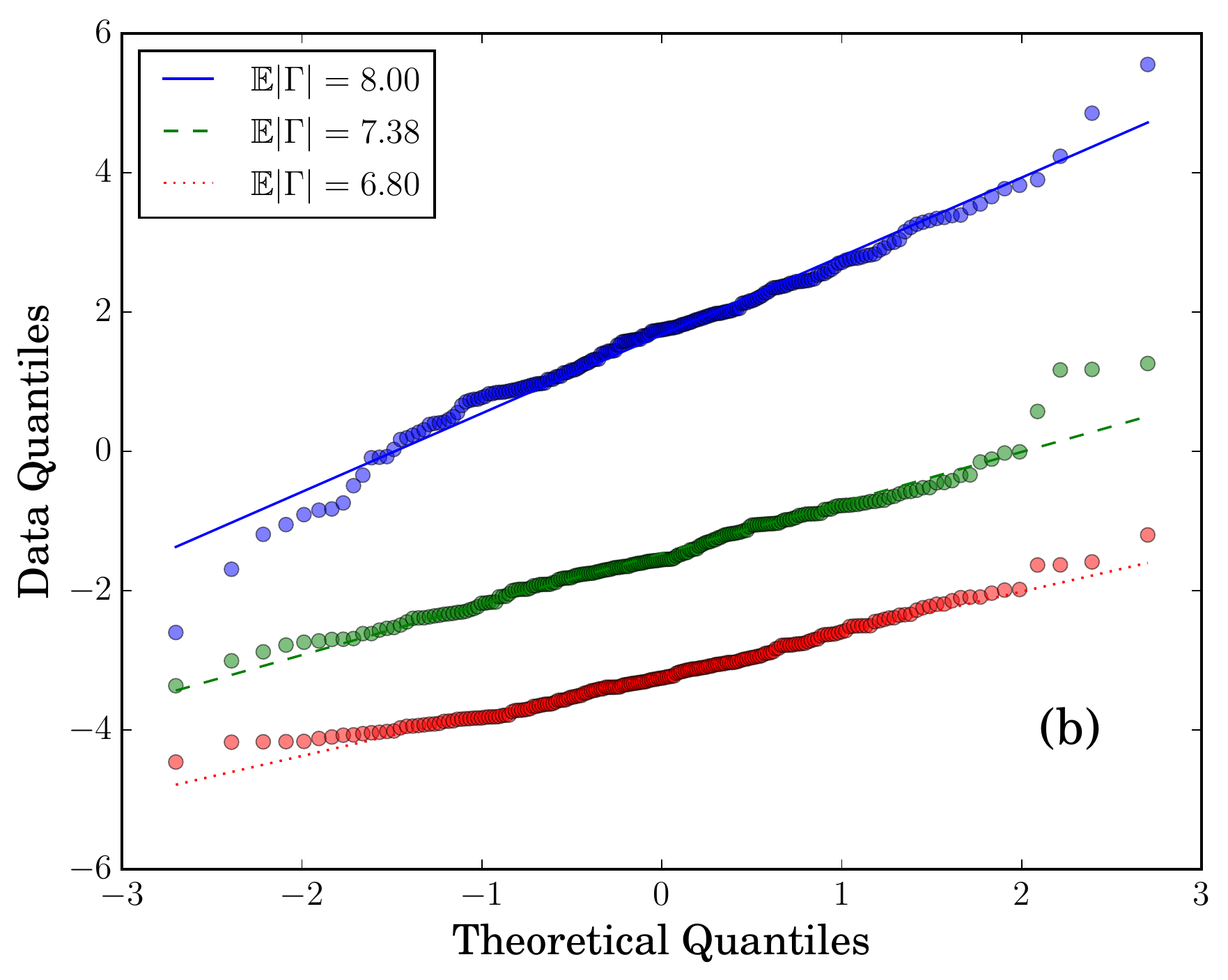}
    \caption{Normal probability plots of ICMs
      $\log_{10} {\rm TTS}_{\textrm{opt}}$ for the instance subclasses
      used to generate the histograms in
      Fig.~\ref{fig:ICMlogTTSOptHistograms}.  A linear relation
      between the Gaussian theoretical and data quantiles implies that
      the data follow a normal distribution. The plots show clearly
      that for the three representative subclasses of
      (a) \instancename{gallus\_26} and (b) \instancename{gallus\_46}
      parametrized by $\Expect{}|\Gamma|$, the histograms are close to
      normal over the majority of their support. In other words,
      ${\rm TTS}_{\textrm{opt}}$ approximately follows a log-normal
      distribution.}
  \label{fig:ICMlogTTSOptQQPlots}
\end{figure*}

In contrast to our results so far, which have considered
instance-averaged difficulty measures,
Fig.~\ref{fig:ICMlogTTSOptHistograms} shows histograms of optimized
ICM $\log_{10} {\rm TTS}$ values for three subclasses of
\instancename{gallus\_26} [Fig.~\ref{fig:ICMlogTTSOptHistograms}(a)]
and of \instancename{gallus\_46}
[Fig.~\ref{fig:ICMlogTTSOptHistograms}(b)] indexed by
$\Expect{}|\Gamma|$. The leftward shift in histogram support shows
clearly that problems tend to become easier with decreasing
$\Expect{}|\Gamma|$. Given the histogram shapes, one may naturally
suspect that ${\rm TTS}_{\textrm{opt}}$ is log-normally distributed.
As $\log {\rm TTS}$ is unlikely to be precisely normal across the
entire data range, we visualize correspondence with a Gaussian via
normal probability plots \cite{chambers:83}, which relate the sample
order statistics, obtained by sorting the data, with the theoretical
means of the corresponding normal order statistics. Deviations from a
linear relation signal lack of Gaussianity. Figure
\ref{fig:ICMlogTTSOptQQPlots} shows the probability plots for the
three subclasses used to generate the preceding histograms for the
\instancename{gallus\_26} [Fig.~\ref{fig:ICMlogTTSOptQQPlots}(a)] and
\instancename{gallus\_46} [Fig.~\ref{fig:ICMlogTTSOptQQPlots}(b)]
classes, respectively.  The relation is close to linear over the
majority of the histogram support, implying in turn that the
${\rm TTS}_{\textrm{opt}}$ is approximately log-normally distributed.
This is consistent with quantities such as the parallel tempering
Monte Carlo autocorrelation time and other roughness measures
\cite{katzgraber:06a,yucesoy:13} having the same property.

While we have argued that $\Expect{}|\Gamma|$ is a good predictor of
mean difficulty, the histogram results show that this value by no means
provides a complete picture. Indeed, when $\Expect{}|\Gamma| = 8$ (i.e.,
$p_6=1$) there is no variance in $|\Gamma_C|$ as all subproblems have
eight-fold degeneracy, yet the $\log {\rm TTS}$ distributions in
Fig.~\ref{fig:ICMlogTTSOptHistograms} nonetheless show considerable
intraclass spread in difficulty. Therefore, there are certainly other
factors at play in predicating the hardness.

\section{Discussion}
\label{sec:discussion}

In this section we discuss generalizations of the planting approach
using lattice animals. Furthermore, we present a case study on how the
approach generalizes to other non hypercubic lattices. Finally, we
discuss the use of the planted problems for fundamental studies of spin
glasses and related statistical systems.

\subsection{Generalization via lattice animals}

\begin{figure}[tb]
  \includegraphics[width=0.85\columnwidth]{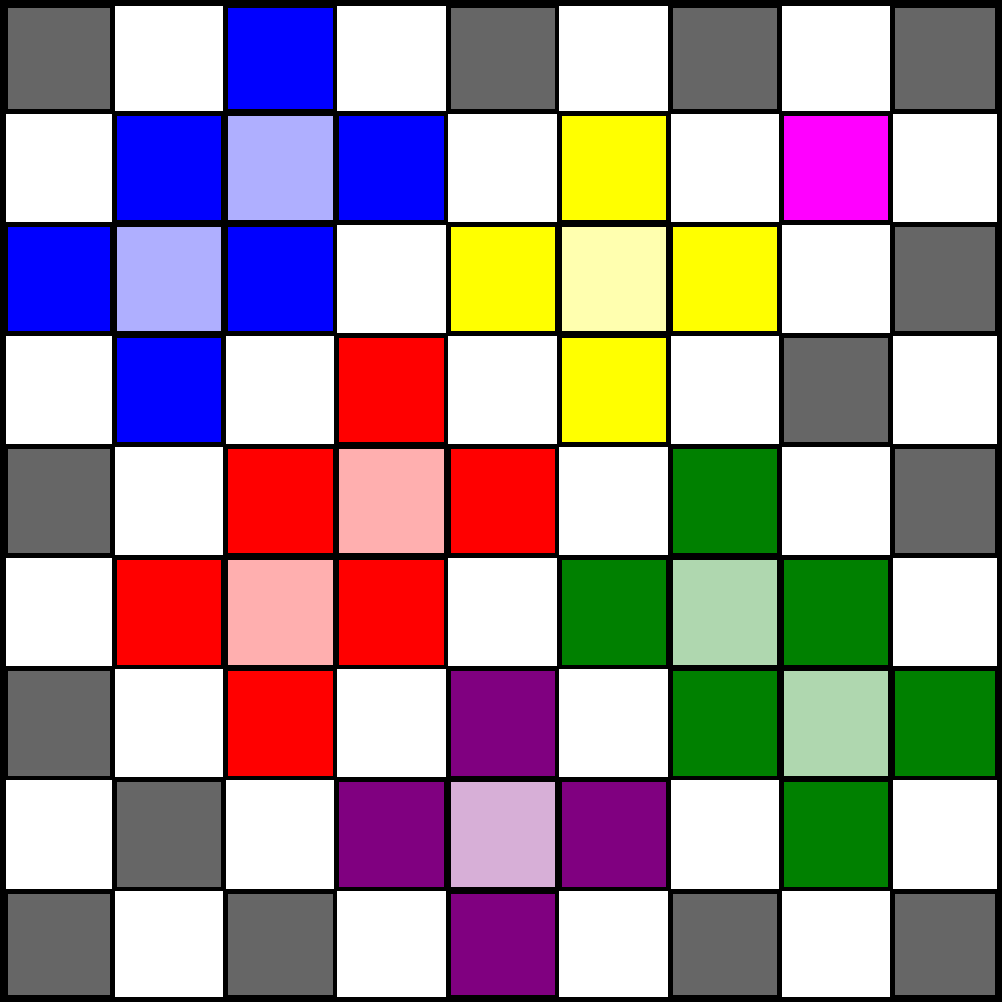}
  \caption{Generalization of unit-cell planting using lattice
    animals, also known as polyominoes. Illustrations are in two
    dimensions for clarity. The analogous procedure in three space
    dimensions is similar.  With lattice animal planting, the
    edge-disjoint subgraphs underlying the subproblems are no longer
    restricted to the unit cells in $\Ccal$ (the checkerboard in two
    space dimensions and the decomposition in
    Fig.~\ref{fig:BCCPartition} for three space dimensions) but are now
    permitted to be connected subgraphs comprised of unions of cells
    from $\Ccal$. As before, subproblem couplers are not added.  Above,
    six such subgraphs in non gray colors are shown, of which only the
    pink one (top right) is comprised of a single cell. If the tree widths of the
    lattice animals are small, their ground states can be computed
    exactly and gauged to the overall target ground state. This
    extension considerably expands types of subproblems that can be
    employed and also introduces an additional mechanism of solution
    hiding via randomization of the employed lattice animals.}
  \label{fig:LatticeAnimals}
\end{figure}

The proposed unit-cell planting technique shows encouraging results
and properties and one may inquire how it may be generalized. In this
section, we present a natural extension of the idea, still assuming
lattice-structured problems, where instead of defining subproblems on
the unit cells of $\Ccal$, they are specified on subgraphs consisting
of their unions. One can verify that such subgraphs, called lattice
animals or polyominoes, also partition the lattice into edge-disjoint
subgraphs, meaning that subproblem couplers are still not added.

A two-dimensional example of decomposition into lattice animals is shown
in Fig.~\ref{fig:LatticeAnimals}. Generalization to three-dimensional
polyominoes obtained by grouping cells from the decomposition shown in
Fig.~\ref{fig:BCCPartition} is straightforward. Shown in non-gray colors
are six lattice animals. Only the pink (top right) one is made of a
single cell. The key difference from the basic method is that unit cells
of a given color are no longer constrained to have their individual
ground states agree; only the complete animal ground state is relevant.
This extension thus considerably extends the types of subproblems that
can be employed and also introduces an additional mechanism of solution
hiding, namely, via randomization of the employed lattice animals, which
would of course be unknown to the would-be adversary. While the tiling
puzzle and CSP interpretation of the problem, suitably modified,
continues to hold in this generalization, under lattice animal
randomization, the adversary would in essence no longer be certain what
puzzle they are even solving.

In this work, we have considered carefully chosen families of
subproblems on three-dimensional unit cells. An exploration of
extensions to more general lattice animals is outside the scope of the
present work. We note, however, that if subproblem couplers are sampled
from a given distribution, then provided the subgraph tree widths
are small, their ground states can be computed exactly \cite{koller:09}
and gauged to the desired overall ground state.

Finally, we note that this lattice animal generalization is also key
when attempting to reduce degeneracy in the planted problems. For
example, by selecting the coupler values from a Sidon set
\cite{sidon:32,katzgraber:15,karimi:17a} of the form
$\{\pm(n-2)/n,\pm(n-1)/n,\pm 1\}$ with, e.g., $n = 50$, the degeneracy is
drastically reduced. Similarly, one could select the couplers from a
distribution of the form
\begin{equation}
|J_{ij}| \in a + (1-a)u ,
\end{equation}
where $u \in [0,1)$ is a uniform random number and $a$ close to $1$
\cite{katzgraber:10a}. However, instead of using basic tiles, more complex
lattice animals must be used to construct the problems.

\subsection{Generalization to arbitrary graphs}
\label{sec:chimera}

The need to benchmark both classical and quantum optimization heuristics
has hastened the development of advanced planting techniques for
solutions of spin-glass-like optimization problems.  We have so far
focused our attention on hypercubic lattice problems. However, quantum
annealing machines typically have hardwired hardware graphs that are
close to planar.  A prototypical example is the chimera graph
\cite{bunyk:14} used by current versions of the D-wave quantum annealing
machines.  When presented with such a situation, one has two possible
options to apply our planting framework. The first is to impose a
lattice structure onto the available graph and the second is to specify
subproblems on altogether different ``unit cells'' than cubes, which of
course, must be graph dependent. We now discuss the first option applied
to the special case of chimera.

\begin{figure}
    \includegraphics[width=0.95\columnwidth]{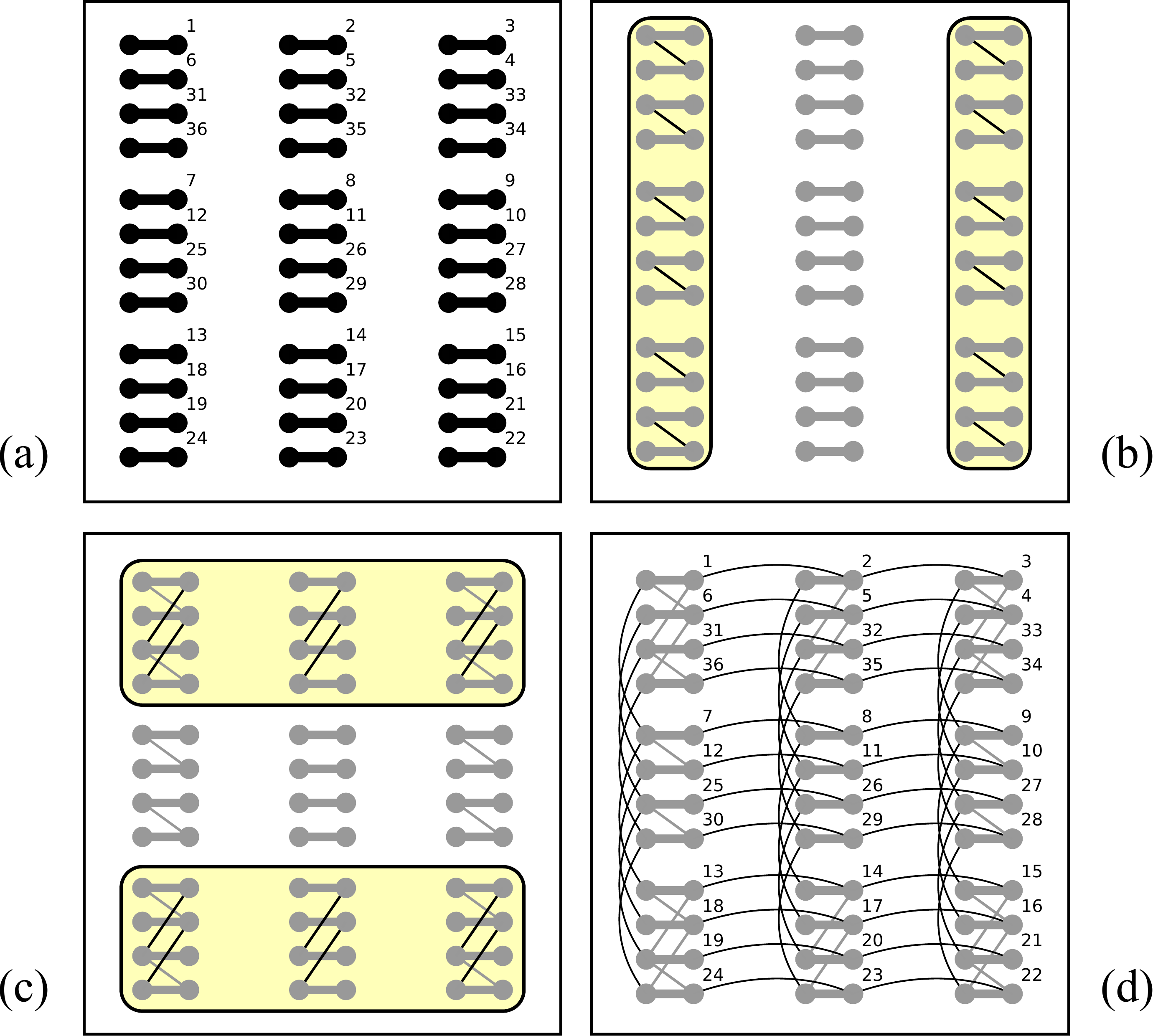}
    \caption{Implementation of a $6\times 6$ square lattice with
      periodic (toroidal) boundary from a $3\times 3$ unit-cell
      chimera graph. (a) Instantiation of two-dimensional logical
      spins from the chimera.  For concreteness, $36$ lattice spins are
      assumed to be labeled in row-fastest order. First, each spin in
      the bipartite cells is paired with the directly opposing spin of
      the same cell. Next, the resultant spin pairs, called dimers,
      are forced to behave as one spin via strong ferromagnetic
      coupling. Each dimer is labeled with the corresponding lattice
      logical spin. (b) In the first and last columns of cells,
      lattice problem edges between dimers $\{1,2\}$ and $\{3,4\}$ of
      each cell are added. Note that only one of the two existing cell
      edges is shown, but if both are used, the coupler strength must
      be suitably divided. (c) In the top and bottom rows of cells,
      lattice edges between dimers $\{1,3\}$ and $\{2,4\}$ of each
      cell are added. Again, only one of the two available edges is
      displayed. (d) All inter cell couplers are specified
      according to the lattice problem.  The lattice variables are
      labeled beside each dimer again, from which one can verify that
      the resultant construction does indeed implement a $6\times 6$
      two-dimensional torus.}
  \label{fig:Chimera2DEmbed}
\end{figure}

Although in principle three-dimensional lattices can be implemented on
the chimera graph \cite{harris:17x}, the required overhead may limit
linear (planted) problem sizes that can be practically studied. On the
other hand, relatively large two-dimensional lattices with periodic
boundaries can be straightforwardly embedded on chimera, with a
relatively modest constant ratio of two chimera variables per lattice
spin. More precisely, a chimera graph of $L\times L$ bipartite unit
cells, comprised of $8L^2$ variables in total, can produce a toroidal
two-dimensional lattice of size $2L\times 2L$. Rather than formally
describe the rather natural procedure, we illustrate it in
Fig.~\ref{fig:Chimera2DEmbed}, where a $6\times 6$ periodic lattice is
created from a $3\times 3$ $K_{4,4}$ cell chimera graph.

So far, we have focused primarily on planting subproblems on
three-dimensional unit cells, in part because planar problems without a
field are solvable in polynomial time \cite{barahona:82}, but as
illustrated in Figs.~\ref{fig:tilingCSPSteps} and
\ref{fig:LatticeAnimals}, the two-dimensional analog is readily
obtained. The analytical tractability of the planar lattice enables a
deep exploration of physical and computational properties, a work which
will be reported elsewhere \cite{perera:18x}. Here we outline the idea
and demonstrate that it does indeed perform well on planar lattices and
nonplanar quasi-two-dimensional chimera graphs using population
annealing simulations.  The two-dimensional subproblems, i.e., the
analogs of the cells shown in Fig.~\ref{fig:FPProblems}, are partitioned
into classes $\{C_i\}$ for $i \in \{1,\ldots,4\}$, within which cells
have $i$ minimizing ground-state configurations each. To achieve the
construction, first define two magnitudes $J_s, J_l > 0$ with $J_l > J_s
$; presently we take $J_l = 2$ and $J_s = 1$. A cell in class $C_i$ is
constructed by setting a random edge to be antiferromagnetic with
magnitude $J_s$, $i-1$ of the remaining edges to be ferromagnetic with
magnitude $J_s$, and the leftover edges to be ferromagnetic with
magnitude $J_l$ \cite{comment:precision}. It is readily verified that
the subproblems do indeed have the specified number of local ground
states which always include the ferromagnetic state.

\begin{figure}
    \includegraphics[width=0.95\columnwidth]{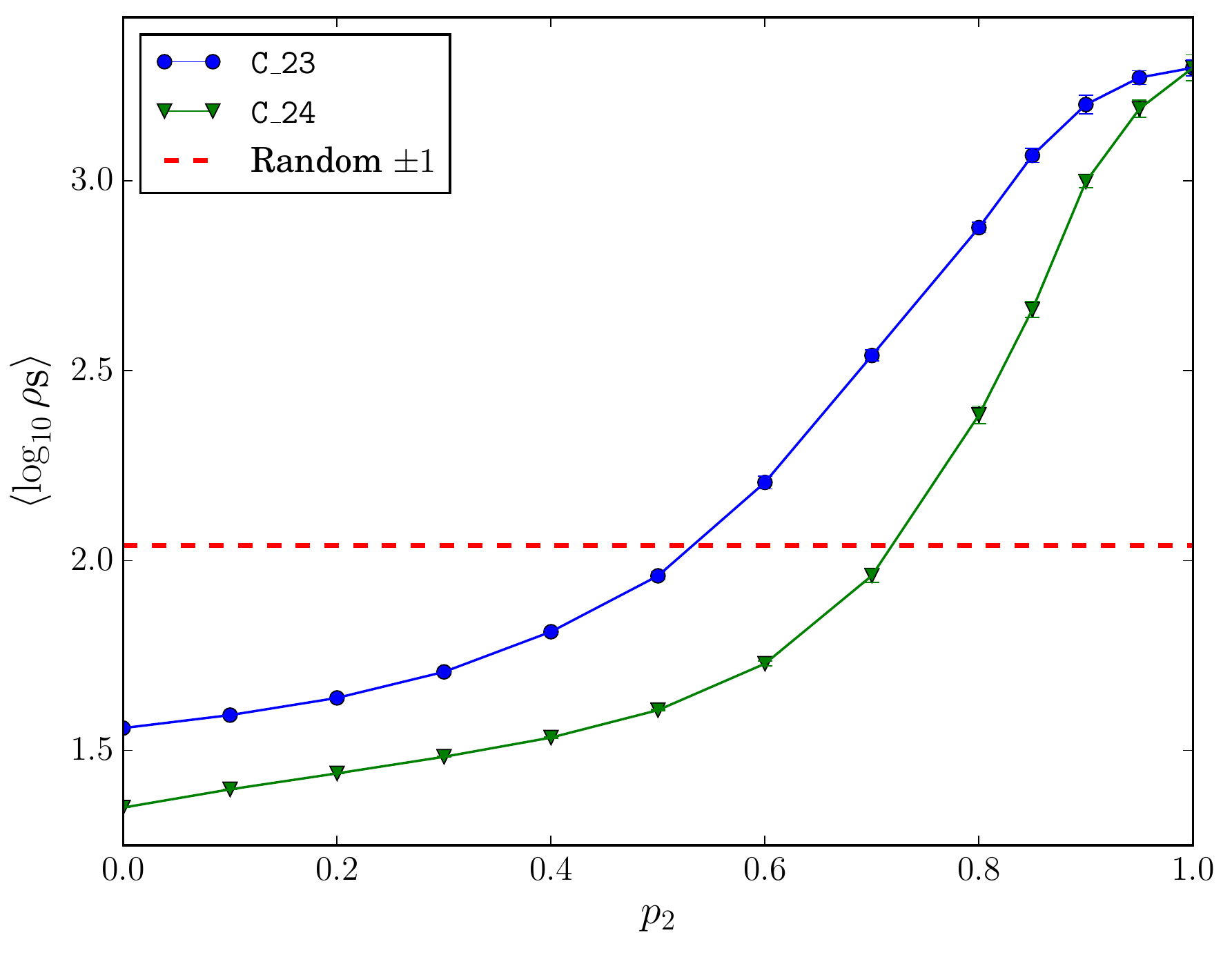}
    \caption{Average log-entropic family size
      $\langle\log_{10} \rho_{\rm S}\rangle$ for $L=24$ periodic planar
      lattice classes \instancename{C\_23} and \instancename{C\_24}.
      Each point corresponds to a subclass in which unit cell
      subproblems are selected from $C_2$ with probability $p_2$, or
      else from $C_3$ for \instancename{C\_23} and from $C_4$ for
      \instancename{C\_24}. Averages are computed over $200$ instances
      from each subclass. As for the study in three dimensions, we see a
      large range of landscape roughness, again with a range of
      subclasses demonstrating greater difficulty than equal-sized
      random bimodal spin-glass problems. The most difficult problems in
      our ensemble are those in which the subproblems are exclusively
      selected from $C_2$, in which the local GS degeneracy is $2$.}
    \label{fig:PAlogRhoP2_2D}
\end{figure}

As was done in three space dimensions, we consider two classes of
problems each comprised of mixtures of two subproblem classes. Problem
class \instancename{C\_24}, consists of mixtures of $C_2$ and $C_4$
cells, while in \instancename{C\_23} the cells may belong to $C_2$ or
$C_3$. Both problem classes are parametrized by $p_2$, the probability
of choosing a cell from $C_2$. The results on $24 \times 24$ periodic
lattices are shown in Fig. ~\ref{fig:PAlogRhoP2_2D}. Again, a wide
range of landscape roughness is seen to be attainable, including a
regime in which the problems are more difficult than random bimodal
($\pm 1$ couplers) spin glasses.  An interesting distinction from
three-dimensional results is that the most difficult problems are not
those in which cells exclusively belong to the maximally degenerate
class $C_4$, but rather the moderately constrained $C_2$ class. In
fact, class $C_4$ is a highly underconstrained regime for this
topology and gives rise to very easy problems. The fascinating
connections between dimension, complexity, and phase behavior are the
subject of ongoing study that goes beyond the scope of this paper.

If one wished to move beyond a regular lattice structure, the natural
objective is a decomposition of the problem graph $G$ into edge-disjoint
subgraphs satisfying some constraint allowing tractable minimization. An
example of a constrained decomposition is into subgraphs with given
minimum vertex degree \cite{yuster:13}. More directly applicable to the
planting context would be a constraint on maximum subgraph tree width
thereby allowing determination of the planted subproblem ground states.
The edge-disjointness property is the key common aspect with the ideas
presented in this paper, as it continues to circumvent the need for
adding subproblem couplers. This idea is clearly a generalization of the
lattice animal methodology discussed previously.  While we have not
presently considered planting using such generic subgraphs, one can
readily envision a heuristic decomposition algorithm that greedily grows
partitioning subgraphs until their tree widths exceed some criterion.

\subsection{Spin-glass physics}

We hope researchers in the field embrace these planted problems to study
physical properties of glassy systems beyond the benchmarking of
optimization heuristics.

Having arbitrarily large planted solutions for hypercubic lattices
allows one to address different problems in the physics of spin glasses.
For example, the computation of defect energies, intimately related to
fundamental properties of these paradigmatic disordered
systems, strongly depends on the knowledge of ground states (see, for
example,
Refs.~\cite{hartmann:97,hartmann:99,palassini:99,palassini:01,hartmann:01b,katzgraber:01}).
Being able to plant problems would drastically reduce the computational
effort in answering these fundamental questions.

Furthermore, by carefully tuning the different instance classes,
problems with different disorder and frustration can be generated. A
systematic study of the interplay between disorder and frustration is
therefore possible for nontrivial lattices beyond hierarchical ones. 
Similarly, being able to tune the fraction of
frustrated plaquettes allows one to carefully study the emergence of
chaotic effects in spin glasses (see, for example,
Refs.~\cite{katzgraber:07,thomas:11e,wang:15a} and references therein).

\subsection{Application-based benchmark problems}

It is well established that random benchmark problems
\cite{ronnow:14a,katzgraber:14} for classical and quantum solvers using
spin-glass-like problems are computationally typically easy.
Furthermore, the control over the hardness of the benchmark problems has
been rather limited either because (post) processing is expensive
\cite{katzgraber:15,marshall:16} or because the benchmark generation
approach does not give the user enough control over the problems to
match, e.g., hardware restrictions \cite{hen:15a}.

Because application-based problems from industrial settings are highly
structured, they pose additional challenges for the vast pool of
optimization techniques designed, in general, for random unstructured
problems. This has sparked the use of problems from industry to generate
hard (and sometimes tunable) benchmark problems. Most notably, the use
of circuit fault diagnosis \cite{perdomo:15,perdomo:17x} has produced
superbly hard benchmarks with small number of variables. However, the
use of application-based problems for benchmarking is in its infancy and
while circuit fault diagnosis shows most promise \cite{perdomo:17x},
many applications have produced benchmark problems that lack the
richness needed to perform systematic studies; see, for example,
Refs.~\cite{perdomo:12,santra:14,rieffel:15,azinovic:17}.

The problems presented in this work are highly tunable and
computationally easy to generate. Furthermore, they can be embedded in
more complex graphs, as is, for example, commonly done on the D-wave
hardware for application benchmarks. Thus, having this tunability not
only should allow for the generation of problems that might elucidate
quantum speedup in analog annealers, but might also help gain a
deeper understanding into quantum annealing for spin glasses in
general. In parallel, having these tunable problems might elucidate
the application scope of specific optimization techniques, both
classical and quantum.

\section{Conclusions}
\label{sec:conc}

We have presented an approach for generating Ising Hamiltonians with
planted ground-state solutions and a tunable complexity based on a
decomposition of the model graph into edge-disjoint subgraphs.  Although
we have performed the construction for three-dimensional cubic lattices
and illustrated the approach with the two-dimensional pendant, the
approach can be generalized to other lattice structures, as shown in
Sec.~\ref{sec:chimera} for the chimera lattice. The construction allows
for a wide range in computational complexity depending on the mix of the
elementary building blocks used.  We corroborated these results with
experiments using different optimization heuristics. Subsequent studies
should discuss constructions with controllable ground-state degeneracy, as
well as the mapping of the complete complexity phase space.

\section*{Acknowledgements}

We acknowledge helpful discussions with Jon Machta, Bill Macready,
Catherine McGeogh, Humberto Munoz-Bauza, and Martin Weigel and are
grateful to Fiona Hanington for thorough suggestions to the
manuscript. F.~H.~would like to thank the Santa Fe Institute for its
hospitality and invitation to a stimulating Working Group on solution
planting, and is indebted to those who introduced him to the original
Eternity puzzle. H.~G.~K.~acknowledges support from the NSF (Grant
No.~DMR-1151387). The Texas A\&M team's research is based upon work
supported by the Office of the Director of National Intelligence
(ODNI), Intelligence Advanced Research Projects Activity (IARPA), via
Interagency Umbrella Agreement No. IA1-1198.  The views and
conclusions contained herein are those of the authors and should not
be interpreted as necessarily representing the official policies or
endorsements, either expressed or implied, of the ODNI, IARPA, or the
U.S.~Government.  The U.S.~Government is authorized to reproduce and
distribute reprints for Governmental purposes notwithstanding any
copyright annotation thereon.  We thank Texas A\&M University and the
Texas Advanced Computing Center at University of Texas at Austin for
providing HPC resources.

\bibliography{comments,refs}

\begin{thebibliography}{66}
\bibitem[{com({\natexlab{a}})}]{comment:free}
\bibinfo{note}{{Referring to the subproblem classes though, a few moments'
  reflection reveals that interestingly, free spin degeneracy is impossible in
  the planted state.}}
\bibitem[{com({\natexlab{c}})}]{comment:p6}
\bibinfo{note}{{Hence, when $p_6=1$, the classes are identical.}}
\bibitem[{com({\natexlab{d}})}]{comment:precision}
\bibinfo{note}{{Hence two bits of precision are required for these classes.}}
\end{thebibliography}

\end{document}